\documentclass[useAMS,usenatbib]{mn2e}
\usepackage{graphicx}
\usepackage{paralist}
\usepackage{amsmath,amssymb,amsfonts,cases}
\newcommand{\echo}{{\it EChO}}
\newcommand{\echosim}{{\it EChOSim}}

\title[Photometric stability of EChO]{Photometric stability analysis of the \\Exoplanet Characterisation Observatory}
\author[I.~P.~Waldmann et al.]
  {I.~P.~Waldmann,$^1$\thanks{Email: ingo@star.ucl.ac.uk}
  E.~Pascale,$^2$  B.~Swinyard,$^{1,3}$ G.~Tinetti,$^1$  A.~Amaral-Rogers$^{2}$
  \newauthor % starts a new line in the
 L.~Spencer,$^{2}$ M.~Tessenyi,$^{1}$ M.~Ollivier$^{5}$ and V. Coud\'e du Foresto$^{6,7}$ \\
  $^1$ Dept. Physics \& Astronomy, University College London, Gower Street, London, WC1E 6BT, UK\\
  $^2$ School of Physics \& Astronomy, Cardiff University, Cardiff, CF24 3AA, UK \\
  $^{3}$ STFC Rutherford Appleton Laboratory, Harwell Oxford, Didcot, OX11 0QX, UK \\
  $^{4}$ Dept. Physics \& Astronomy, University of Leicester, Leicester, LE1 7RH, UK \\
  $^{5}$ Institut d'Astrophysique Spatiale, B‰timent 121, Universit\'e de Paris-Sud, 91405 ORSAY Cedex, France\\
  $^{6}$ Observatoire de Paris (LESIA), 5 place Jules Janssen, F-92190 Meudon, France \\
  $^{7}$ Center for Space and Habitability, University of Bern, Sidlerstrasse 5, CH-3012, Bern, Switzerland 
  }
\begin{document}

\date{}

%\pagerange{\pageref{firstpage}--\pageref{lastpage}} \pubyear{2002}

\maketitle

\label{firstpage}

\begin{abstract}
Photometric stability is a key requirement for time-resolved spectroscopic observations of transiting extrasolar planets. In the context of the Exoplanet Characterisation Observatory (\echo) mission design, we here present and investigate means of translating space-craft pointing instabilities as well as temperature fluctuation of its optical chain into an overall error budget of the exoplanetary spectrum to be retrieved. Given the instrument specifications as of date, we investigate the magnitudes of these photometric instabilities in the context of simulated observations of the exoplanet HD189733b secondary eclipse. 
\end{abstract}


\section{Introduction}

The last decade has seen a surge in exoplanetary discoveries, with $\sim$ 850 planets confirmed \citep{schneider11} and over two thousand Kepler candidates \citep{borucki11,batalha13} waiting for confirmation.
With the precise measurements of  their masses and radii we have gained a staggering wealth of information for a plethora of targets and planetary types. In this large and growing consensus of foreign worlds, some afford us the opportunity of further characterisation. By studying the exoplanets' atmospheres, we can not only infer their chemical make-up but also constrain their climates, thermodynamical processes and formation histories. 
The sum total of this knowledge will allow us to understand planetary science and our own solar system in the context of a much larger picture.

Using {\it Hubble}, {\it Spitzer} as well as ground based facilities, the success of exoplanetary spectroscopy of transiting \citep[e.g.][]{beaulieu10, beaulieu11, berta12, charbonneau02,charbonneau08, deming05, deming07, grillmair07, grillmair08, richardson07, snellen08, snellen10a, snellen10b,  bean11b,  stevenson10, swain08, swain09a,swain09b,tinetti07a, tinetti07b, tinetti10, thatte10, mandell11, sing09, sing11, pont08, burke07, burke10, desert11, redfield08, waldmann12, waldmann13, crouzet12,brogi12} as well as non-transiting planets \citep[e.g.][]{janson10, currie11} has been remarkable in recent years. 

Following these recent successes and in the frame of ESA's Cosmic Vision programme, the Exoplanet Characterisation Observatory ({\it EChO}) has been considered as medium-sized M3 mission candidate for launch in the 2022 - 2024 timeframe \citep{tinetti12b}. 
The current `Phase-A study' space-mission concept is a 1.2 metre class telescope, passively cooled to $\sim$50 K and orbiting around the second Lagrangian Point (L2). The current baseline for the payload consists of four integrated spectrographs providing continuous spectral coverage from 0.5 - 16$\mu$m at resolution ranging from R $\sim$ 300 to 30. For a detailed description of the telescope and current payload design studied by our instrument consortium, we refer the reader to the literature \citep{puig12,tinetti12, tinetti12b, swinyard12, eccleston12, reess12, adriani12, zapata12, pascale12, focardi12, tessenyi12}.  
In order to observe the spectrum of the extrasolar planet, the \echo~mission uses the time-resolved spectrophotometry method to observe transiting exoplanets. The spectrum of the planet is either seen in transmission when the planet transits in front of the star along our line of sight or in emission when the planet's thermal contribution is lost during secondary eclipse when the planet disappears behind its host.  

By taking consecutive short observations we follow the transit or secondary eclipse event of an extrasolar planet. This approach results in an exoplanetary lightcurve for each individual wavelength channel. The spectrum of the planet is then derived by model fitting said lightcurves and recording the eclipse depths at each wavelength bin. The amplitudes of these modulations on the mean of the exoplanetary lightcurve depths are unsurprisingly small. Taking the case of the secondary eclipse emission spectrum we typically find the planetary contrast to be of the order of 10$^{-3}$ for hot-Juptiers in an orbit around a typical K0 star, and a much less favourable 10$^{-5}$ contrast for so called `temperate Super-Earths' orbiting M-dwarf stars \citep{tessenyi12}. A mission such as \echo~hence needs to provide a sensitivity that either matches or exceeds these contrasts over the duration of a transit/eclipse observation lasting one (e.g. GJ1214b, \citealt{charbonneau09}) to tens of hours (e.g. HD80606b, \citealt{fossey09}). 
The required instrument stability over said time spans can be achieved by providing a high photometric stability.
There are three main factors that may introduce photometric variability over time and may limit the photometric stability. These are: 

\begin{enumerate}
\item {\it Pointing stability of the telescope:}\\ The 1$\sigma$ pointing jitter of the satellite is currently base-lined to be of the order of 10 milli-arcsec from 90s to 10h of continuous observation\footnote{ESA report: EIDA-R-0470}. Effectively this defines the maximum Performance Reproducibility Error (PRE) over ten hours, specifying the reproducibility of the experiment. These pointing drifts manifest themselves in the observed data product via two mechanisms: 1) the drifting of the spectrum along the spectral axis of the detector, from here on referred to as `spectral jitter'; 2) the drift of the spectrum along the spatial direction (or `spatial jitter'). The effect of pointing jitter on the observed time series manifests as non-Gaussian noise correlated among all detectors in all focal planes of the payload and is characterised by the power-spectrum of the telescope pointing. The amplitude of the resultant photometric scatter depends on the pointing jitter power spectrum, the PSF of the instruments, the detector intra-pixel response and the amplitude of the inter- pixel variations.

\item {\it Thermal stability of the optical-bench and mirrors:}\\ 
Thermal emission of the instrument, baselined at 45K, is a source of photon noise and does hence not directly contribute to the photometric stability budge beyond an achievable signal to noise ratio (SNR) of the observation. However fluctuations in thermal emissions constitute a source of correlated noise in the observations and need to be maintained at amplitudes small compared to the science signal observed. Given a $\sim$45K black-body peaks in the far-IR, we are dominated by the Wien tail of the black body distribution, resulting in steep temperature gradients and stringent requirements on thermal stability in the long wavelength instruments of the \echo~payload. Additionally to variable photon noise contribution, thermal variations may impact dark currents and responsivities of the detector which must be taken into account. 

\item {\it Stellar noise and other temporal noise sources:}\\ Whilst beyond the control of the instrument design, stellar noise is an important source of temporal instability in exoplanetary time series measurements \citep{ballerini11}. This is particularly true for M dwarf host stars as well as many non-main sequence stars. Correction mechanisms of said fluctuations must and will be an integral part of the \echo~science study but goes beyond the instrument photometric stability discussion presented here.
\end{enumerate}

 Along with the achievable SNR, the photometric stability of the instrument is the deciding factor for the success of missions or facilities aiming at transit spectroscopy. 

In this paper we study the effects of the spatial/spectral jitter and thermal variability on the photometric stability budget of the instrument and telescope.
%We will use the end-to-end mission simulator \echosim~ (Pascale et al., in prep.) to test the impact of these stability requirements on the SNR of the `observed' spectra of selected key cases. 

\section{EChO and EChOSim}

Given \echo~is currently in its Phase-A study phase, we concede that a noise budget and stability analysis, such as this one, can only be preliminary. Here and in Pascale et al. (in prep.), we present methodology used for the testing and optimisation of the current instrument design. The study presented in this paper draws on the end-to-end mission simulator \echosim, which is discussed in detail in Pascale et al. (in prep.). 

In the simulations of this paper we assume \echo~to have a 1.2 metre diameter primary mirror off-axis telescope. The light beam is simultaneously fed into five spectrographs via dichroic beam splitters: Visible (Vis), short wavelength IR (SWIR), mid-wavelength IR (MWIR1 and MWIR2) and long-wavelength IR (LWIR) instruments covering the spectral range from 0.5 - 16$\mu$m. 

In order to provide an end-to-end simulation of the observation, \echosim~incorporates a full simulation of the science payload including the telescope, as well as the astrophysical scene including Zodi emissions.
\echosim~uses realistic mirror reflectivities and estimates the instrument transmission function for each channel as a function of wavelength. This includes transmission, optical throughput and spatial modulation transfer function. Using tabulated emissivity values as a function of wavelength, \echosim~also estimates the thermal emission spectrum of the several optical elements of the telescope.
The transmission through the dichroic chain is simulated, and the incoming radiation is split among the 5 instruments assuming realistic transmittance and reflectivity data (Pascale et al., in prep.). 
Dispersion and detection by the focal plane array are simulated. The dispersion of the light over the focal plane is modelled by a linear dispersion law, which is related to the sampled spectral resolving power $R$.

The full focal plane illumination due to the backgrounds and the science signal's point spread function (PSF) is calculated and convolved with realistic intra-pixel variations (see section~\ref{sec:spatial}). 
Photon noise, read noise and pointing jitter noise are calculated on a pixel by pixel basis and time series of the observation are generated. These time series are then analysed and model fitted using \citet{mandel02} analytical solutions and an adaptive Metropolis-Hastings Markov Chain Monte Carlo (MCMC) algorithm \citep{haario01,haario06,hastings70}. For a more detailed list of the instrument parameters assumed in these simulations see table~\ref{inputtable} and Pascale et al. (in prep.).

\begin{table*}
\center
\caption{Simulation parameters for \echosim} \label{inputtable}
\begin{tabular}{l | c c c c c}
\hline
Parameters & Vis & SWIR & MWIR1 & MWIR2 & LWIR \\
  \hline
Wavelength range ($\mu$m) & 0.5 - 2.0 & 2.0 - 5.0 & 5.0 - 8.5 & 8.5 - 11.0  & 11.0 - 16.0\\
Resolution ($R$) & 330 & 490 & 35 & 40 & 40 \\
Linear Dispersion ($\mu$m/$\mu$m) & 15840 & 6000 & 689 & 374 & 222 \\
Pixel size ($\mu$m) & 90 & 15 & 30 & 30 & 30 \\
Detector size (pixels) & 256$\times$10 & 1111$\times$18 & 63$\times$10 & 40$\times$10 & 50$\times$10 \\
Slit width (pixels) & 2 & 2 & 2 & 2 & 2\\
Effective focal number (F$_{\#}$) & 4.0 & 2.08 & 1.26 &1.0 & 2.0 \\
Effective focal length (F$_{eff}$,nm) & 5150 & 2680 & 1620 & 1290 & 2600 \\
Aberration parameters ($K_{x},K_{y}$) & 2.5483590431 & 1.4383968346 & 0.9256307428 & 0.6923626555 &1.3561429817 \\
Dichroic emission ($\%$) & 3.0  & 3.0  & 3.0  & 3.0  & 3.0\\
 \end{tabular}
\end{table*}

\section{Frequency bands of interest}

\begin{figure}
\center
\includegraphics[width=8cm]{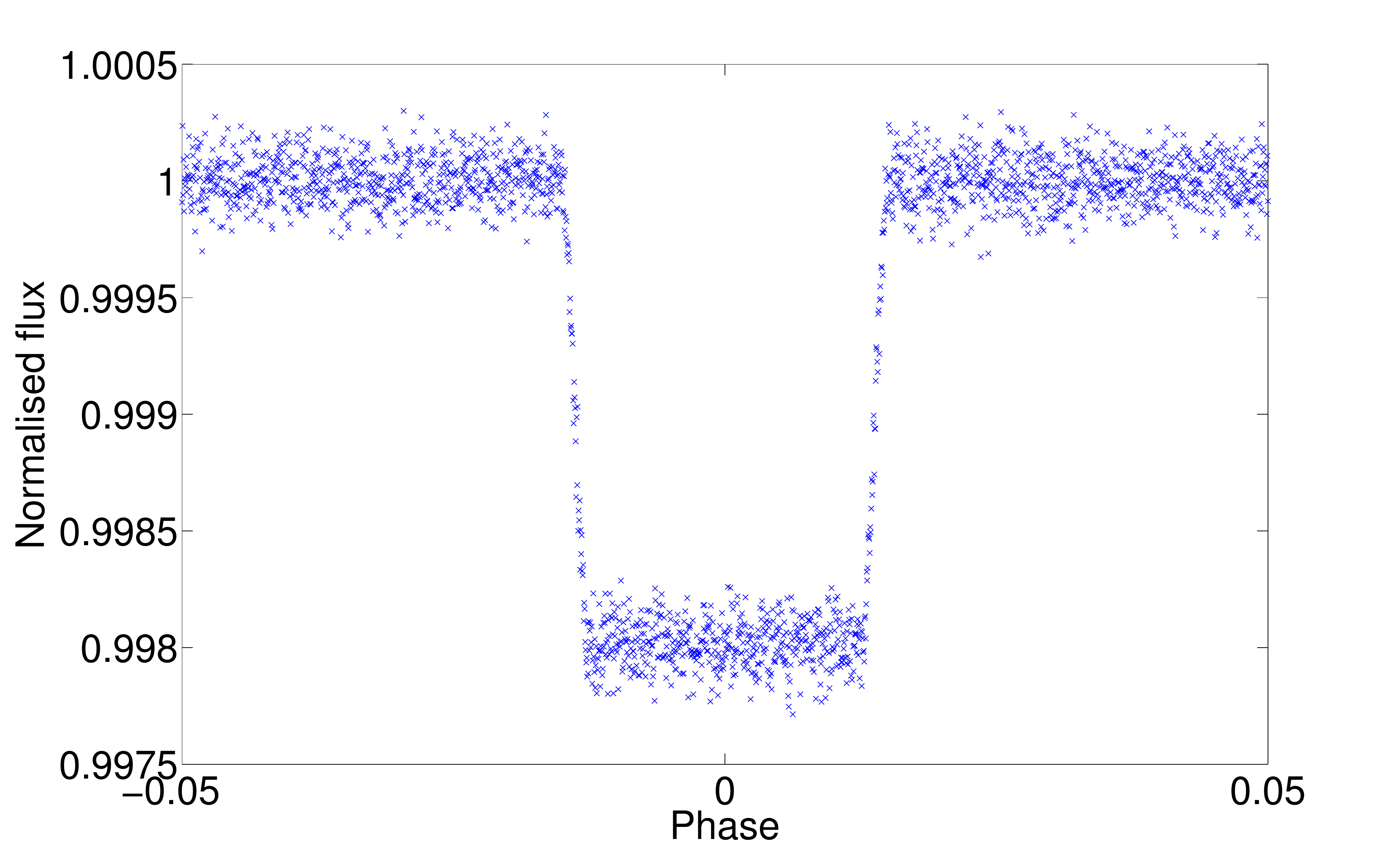}
\caption{Secondary eclipse lightcurve of a hot-Jupiter type exoplanet with eclipse duration of 720min \citep{mandel02}. Noise at 10$^{-4}$ level was added. }
\label{frequ1}
\end{figure}

\begin{figure}
\center
\includegraphics[width=8cm]{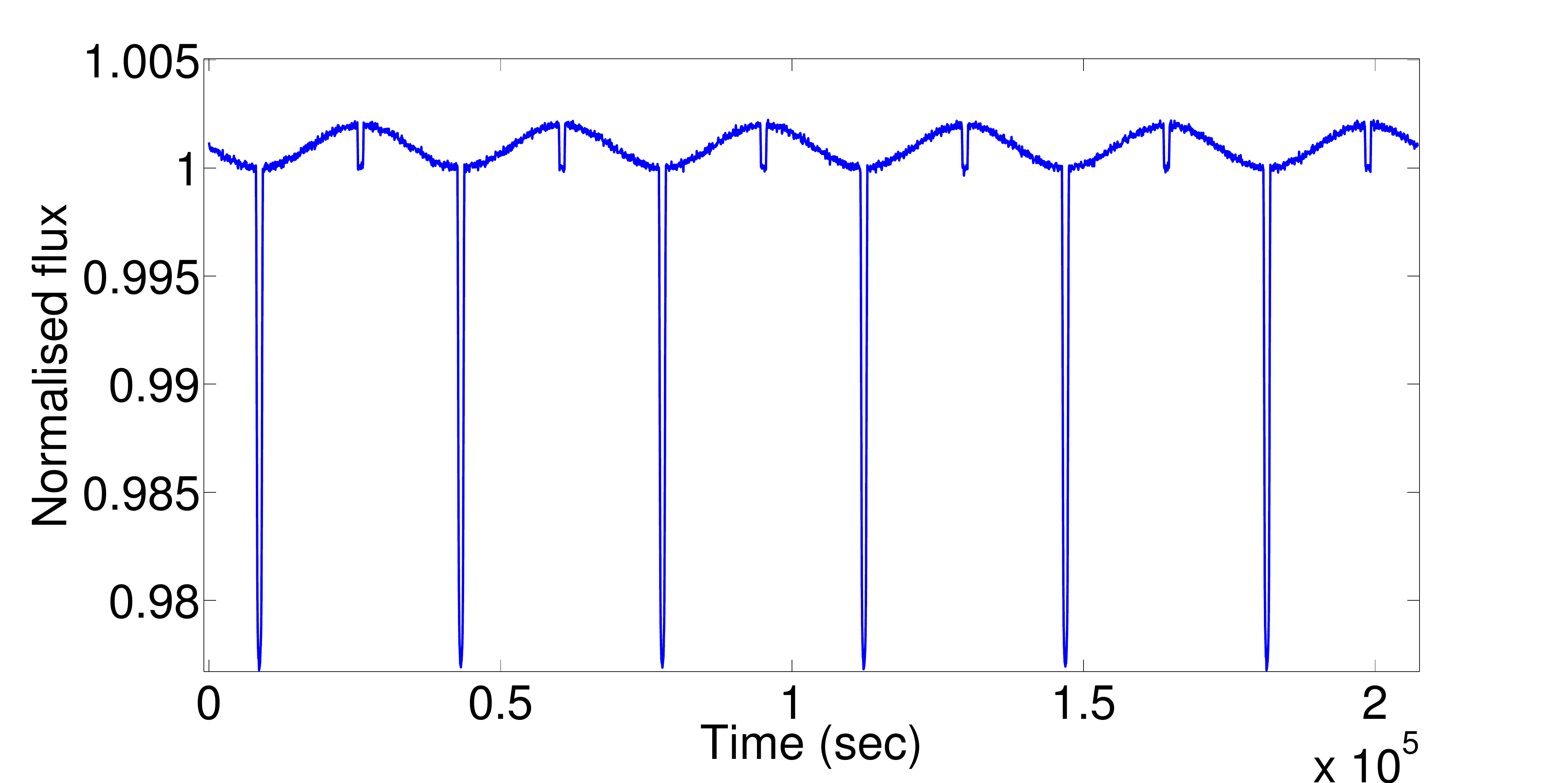}
\caption{Time series of 6 orbits of a hot-Jupiter (akin to HD189733b, \citealt{torres08}). The deep troughs are limb-darkened transits \citep{mandel02,claret00}, smaller troughs are secondary eclipses and sinusoidal variations are due to the planetary phase curve as the planetary day-side rotates in and out of view. White noise of the level of 10$^{-4}$ was added.}
\label{frequ2}
\end{figure}

\begin{figure}
\center
\includegraphics[width=8cm]{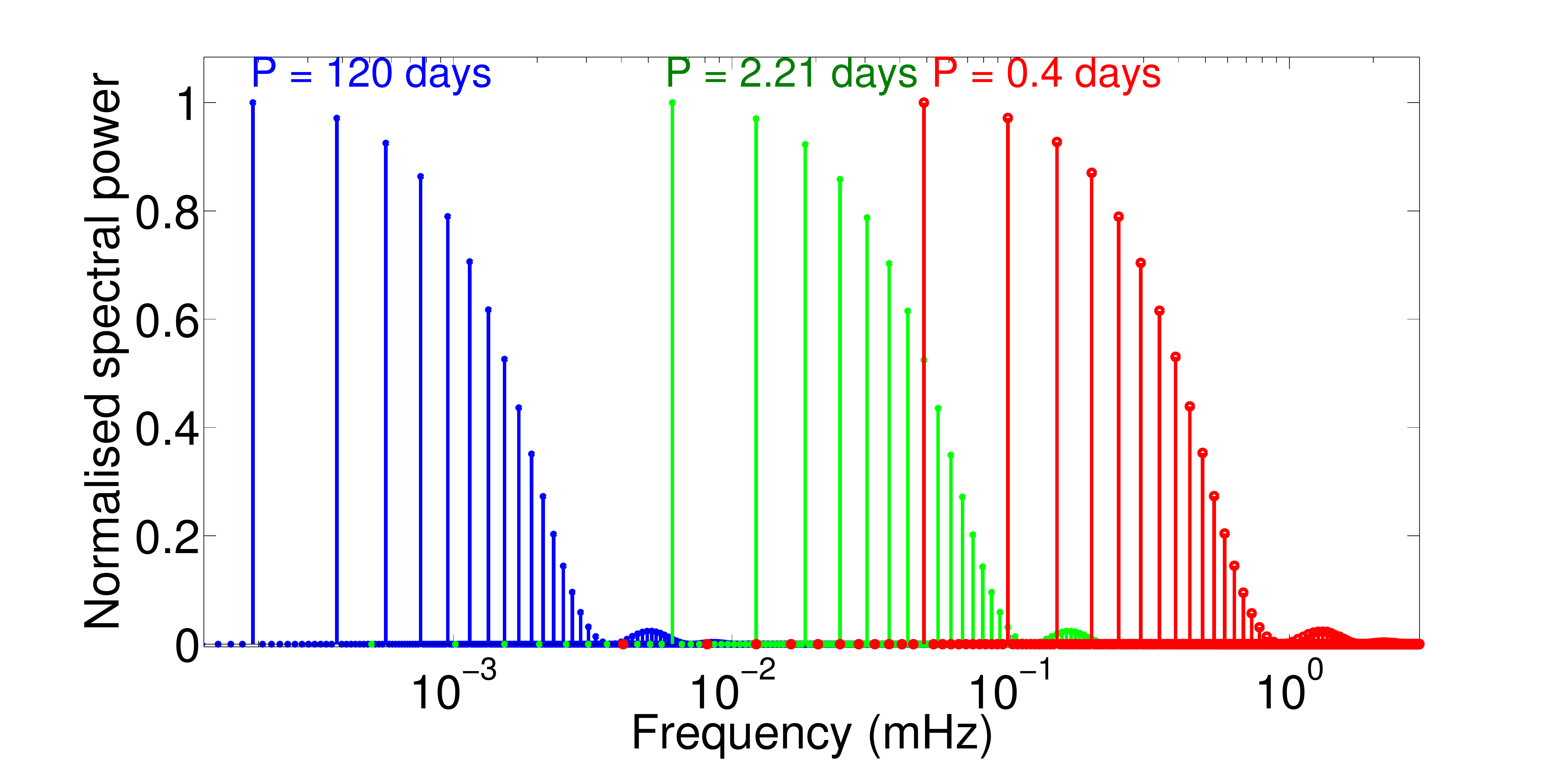}
\caption{Power spectra of time series shown in figure~\ref{frequ2} for different orbital periods. Blue: Period = 120 days, Green = 2.21 days (akin to HD189733b), Red = 0.4 days. The sensitive frequency range extends from 1.9x10$^{-4}$ - 1.7x10$^{-3}$ Hz.}
\label{frequ3}
\end{figure}

 Figure~\ref{frequ1} shows an example of a secondary eclipse of a typical hot-Jupiter planet. Figure~\ref{frequ2} shows the signal observed by \echo~over the duration of 6 planetary orbits of a hot-Jupiter. From these figures it can easily be seen that time-correlated noise has the greatest impact on the retrieved science at temporal variation frequencies comparable to those of the transit/eclipse event, or multiples thereof. Figure~\ref{frequ3} shows the frequency domain representation of Figure~\ref{frequ2} given a variety of orbital periods. Here the desired signal is contained in discrete frequencies and their respective overtones. Frequency ranges beyond these shown in Figure~\ref{frequ3} can safely be filtered out using pass-band filters, or normalisation using low order polynomials in the time domain, without impairing the transit morphology. Given the range of transit periods observed and the goal of accurate ingress and egress mapping, we find the ``crucial frequency band'' to be from 1.9x10$^{-4}$ to ~1.7x10$^{-3}$ Hz, outside of which slow moving trends and high-frequency noise can effectively be filtered. This approach of slow moving trend removal is well tested for Kepler data \citep[e.g.][]{gilliland10}. The overall critical frequency band for \echo~is determined by the longest observation expected and the need to Nyquist sample the highest expected frequencies. We hence limit our simulations of pointing jitter or temperature fluctuations to this frequency range of interest.

\section{Spectral Jitter}
\label{sec:spectral}

As described in the introduction, the telescope pointing jitter has the effect of translating the spectrum on the detector along both, the spatial and spectral directions. Whilst the real translation is a combination of these orthogonal components, we will consider the effect of both these translations independently of each other. With the observing approach of taking consecutive spectra over a given length of time, we obtain a flux time series for each pixel or resolution element $\Delta \lambda$. Should, during the course of the observation, the stellar spectra be shifted along the detector the same pixel will not observe the same stellar flux but that corresponding to the shifted stellar spectrum. If we are to construct a time series for each resolution element of the detector, we consequently need to re-sample the stellar spectra to a common grid. The ability to re-sample the individual stellar spectra to a uniform wavelength grid depends on how well we can determine these relative shifts.  

In this section we will simulate a series of consecutively observed stellar spectra and shift each one of them according to a pointing jitter distribution derived from the {\it Herschel} Space Telescope. We will fit stellar absorption lines of these spectra and use the derived centroids to determine the accuracy with which we can determine the spectral drift. This is translated into a final post-correction flux error. 
It is worth noting that inter and intra pixel variations are less critical in this case as the the Nyquist sampling adopted at instrument level spreads each resolution element over at least two adjacent pixels.

\subsection{Synthesising pointing jitter from Herschel observations}

\begin{figure}
\center
\includegraphics[width=9cm]{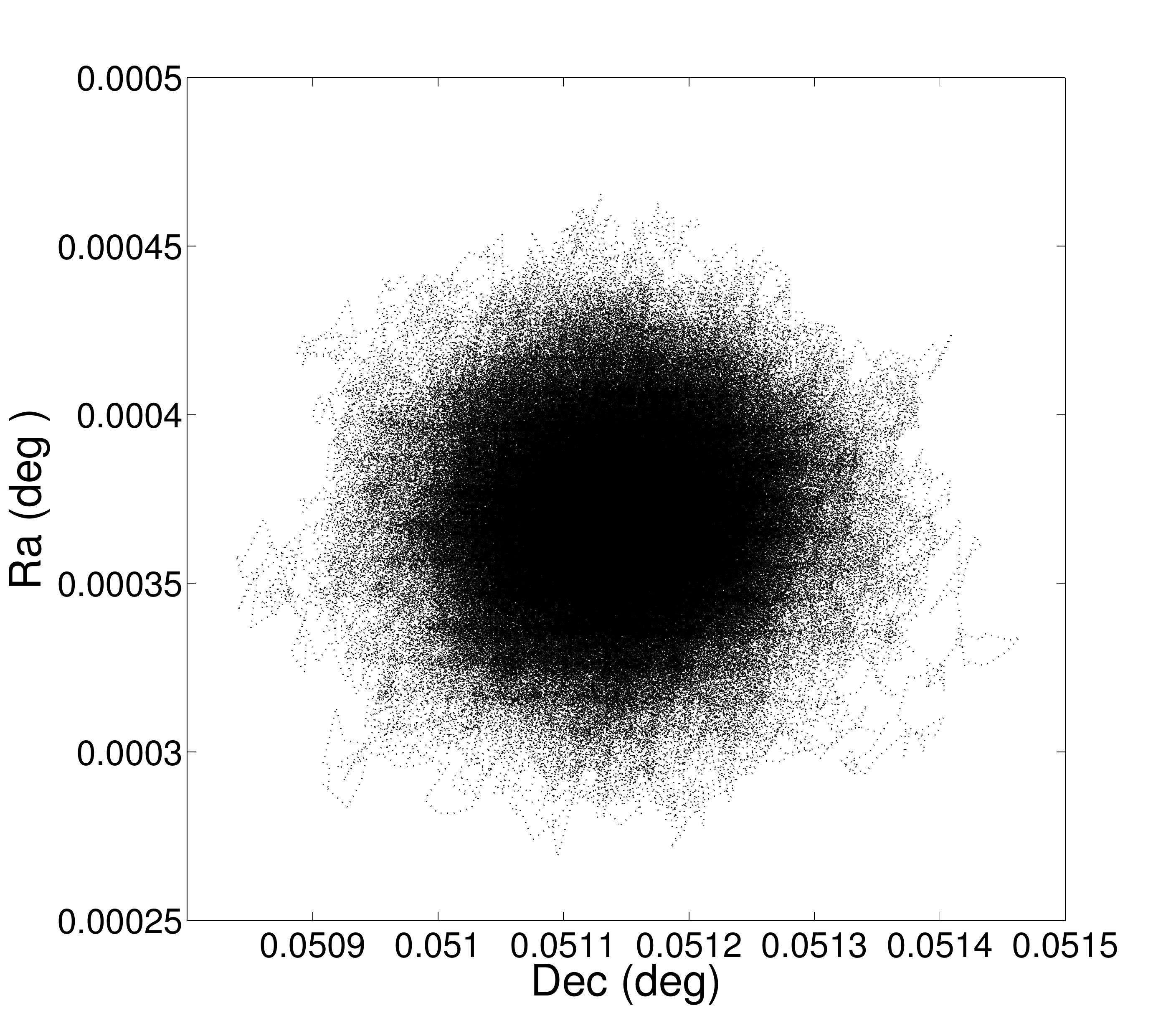}
\caption{Herschel pointing over a three hour continuous observation.}
\label{herschelpoint}
\end{figure}

In order to simulate the pointing jitter of the \echo~mission as realistically as possible, we synthesised the pointing patterns of the {\it Herschel} Space Telescope for which real pointing information exists (Swinyard private communication). Figure~\ref{herschelpoint} shows Herschel pointings recorded for a three hour time period with a sampling frequency of $\sim$1 second. The pointing jitter distance for Herschel, $\Delta_{Her}$, is given by the Pythagorean argument relative to their means:

\begin{equation}
\Delta_{Her} = \sqrt{Ra^{2} + Dec^{2}}.
\end{equation}

Assuming the pointing jitter in $Ra$ and $Dec$ are normally (Gaussian) distributed, we can state that the probability distribution function (pdf) of $\Delta_{Her}$, $P(\Delta_{Her})$, is given by a Rayleigh distribution. The Rayleigh distribution is the pythagorean argument of two orthogonal Gaussian distributions. However, for a real system such as Herschel, the pointing jitter is described by a Gaussian and a non-Gaussian component. These non-Gaussian components propagate to $P(\Delta_{Her})$ in the form of an increased skew and broader wings than those of a pure Rayleigh distribution. We can approximate $P(\Delta_{Her})$ using the more general Weibull distribution of which the Rayleigh distribution is a special case. The Weibull distribution interpolates between a Rayleigh and an Exponential distribution and is hence ideally suited to describe the broader wings and skew of the {\it Herschel} pointing jitter pdf. The Weibull distribution is given by 

\begin{equation}
\large
P_{wbl}(x; \tau, \kappa) =  \left\{ 
  \begin{array}{l l}
    \frac{\tau}{\kappa} (\frac{x}{\kappa})^{\kappa-1} e^{-(x/\tau)^{\kappa}} & \quad x \geq 0,\\ 
    0 & \quad x < 0,\\
  \end{array} \right.
  \label{equwbl}
\end{equation}

\noindent where $\tau$ is known as the amplitude coefficient and $\kappa$ as the distribution shape coefficient. Equation~\ref{equwbl} reduces to an Exponential and a Rayleigh distribution for $\kappa = 1$ and $\kappa = 2$ respectively. The mean and the variance of the Weibull distribution are given by:

\begin{equation}
\mu_{wbl} = \tau \Gamma(1+ 1/\tau)
\label{equwblmean}
\end{equation}

\begin{equation}
\sigma_{wbl}^{2} = \tau^{2} \Gamma(1+2/\tau) - \mu^{2}
\label{equationwblvar}
\end{equation}

\noindent where $\Gamma$ is the gamma function \citep{riley02}. 

\subsubsection{Scaling the Herschel pointing distribution}

\begin{figure}
\center
\includegraphics[width=8cm]{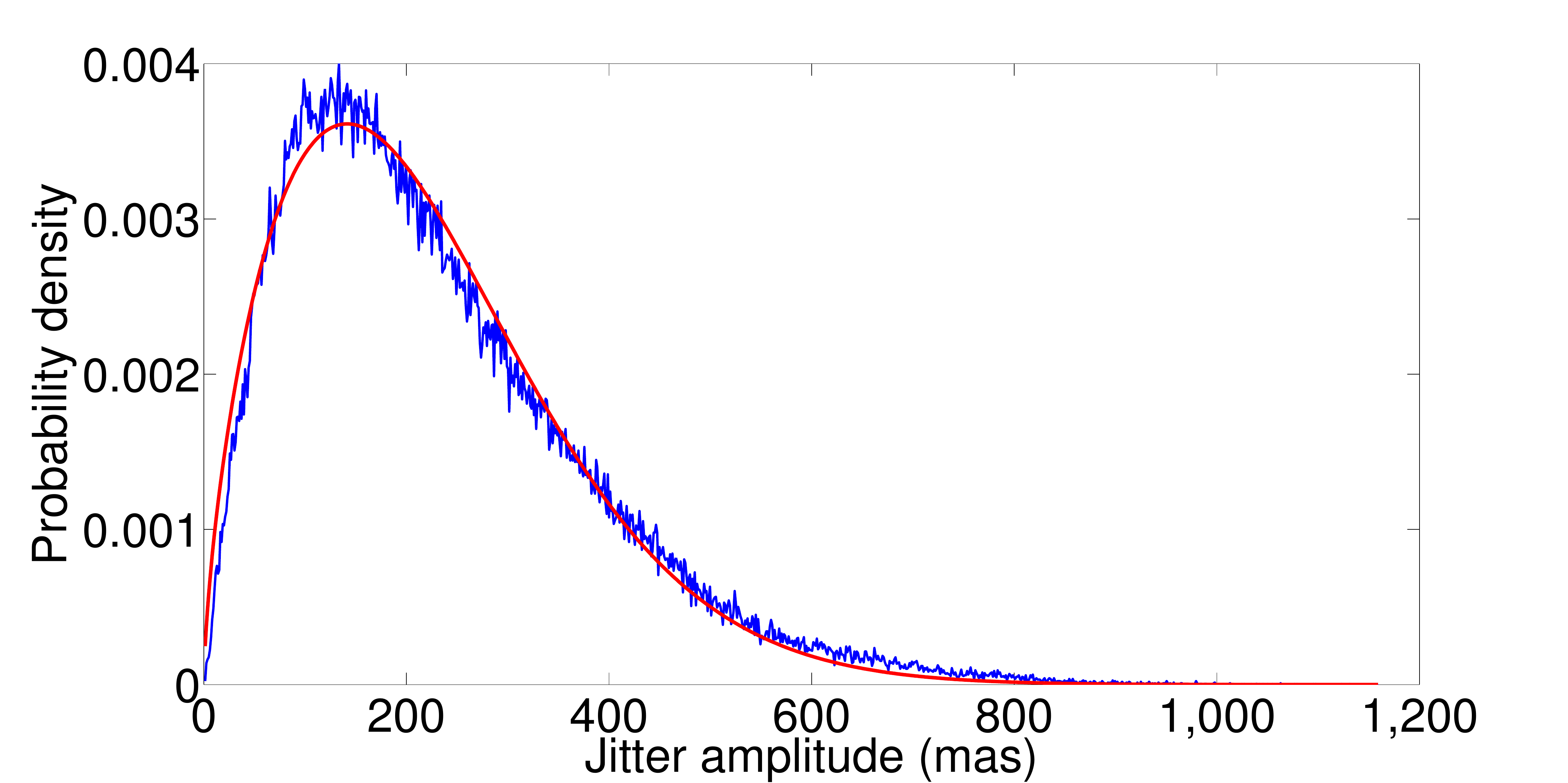}
\caption{Probability distribution function of the pointing jitter length of Herschel}
\label{herschelpdf}
\end{figure}

\begin{figure}
\center
\includegraphics[width=8cm]{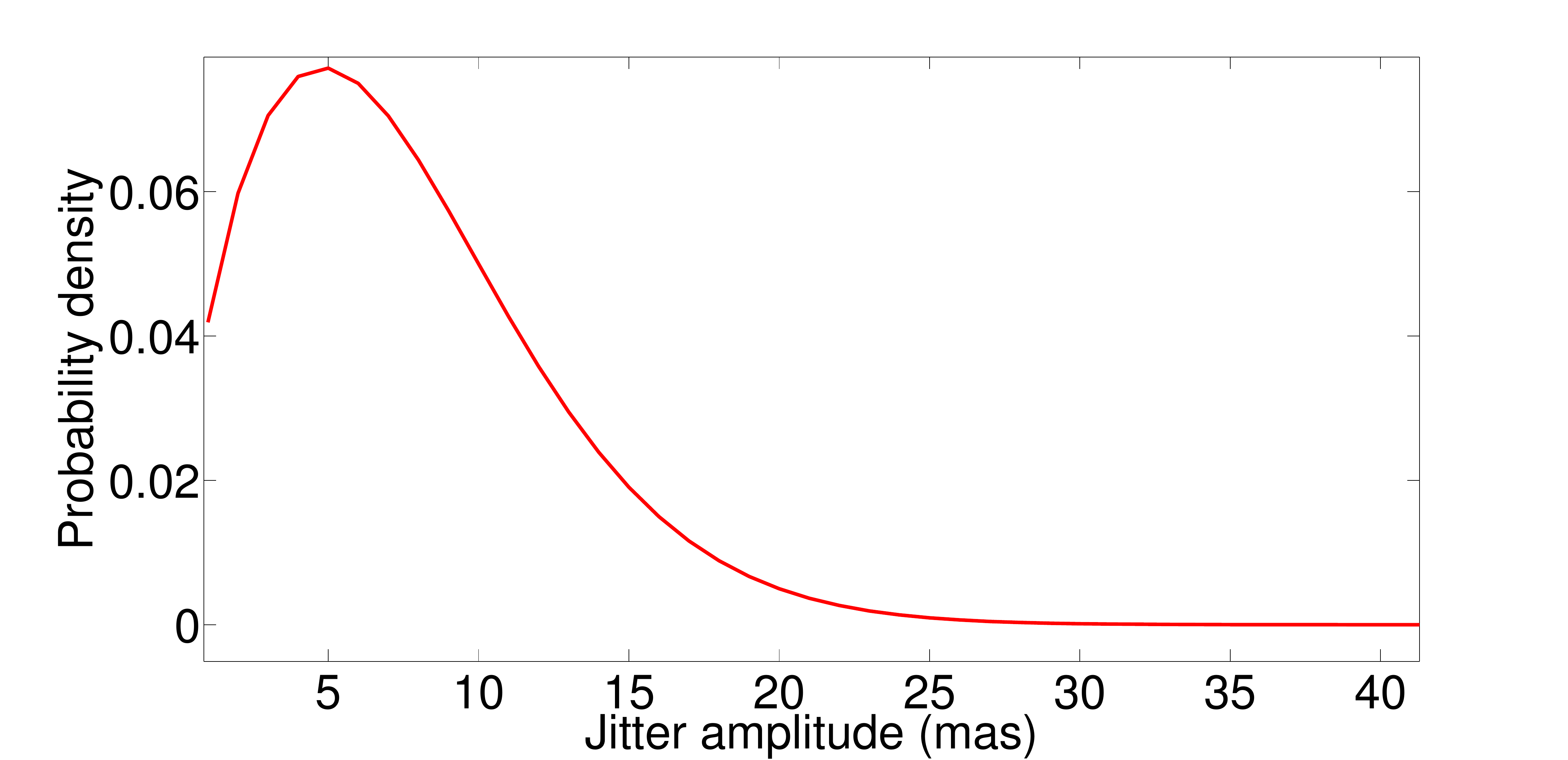}
\caption{Synthesised probability distribution for EChO pointing jitter. }
\label{echopdf}
\end{figure}

Figure~\ref{herschelpdf} shows $P(\Delta_{Her})$ (blue) obtained from the pointing information in figure~\ref{herschelpoint} and the best fitting Weibull distribution, $P_{wbl}(\cdot)$, with $\tau = 247.71$ and $\kappa = 1.6547$. The mean of the distribution is $\mu = 221.45$~mas. In order to obtain the scaled distribution to a jitter amplitude of 10~mas predicted for EChO, we maintain the shape parameter at $\kappa = 1.6547$ and re-derive the scaling parameter, $\tau$, for $\mu = 10$~mas. This yields the pointing jitter distribution $P(\Delta_{Echo})$ shown in figure~\ref{echopdf}. \\

We now randomly sample from $P(\Delta_{Echo})$ to obtain $\Delta_{Echo}$ for $M$ number of observations in our simulated observing run. Assuming a resolving power of $R = 300$ for the NIR channel (1.0-2.5$\mu$m), we can calculate the spectral resolving power to be 0.0159nm/mas. Hence, we can express $\Delta_{Echo}$ as function of spectral wavelength drift  and from hence forth denote the \echo~jitter as $\Delta_{m}(\lambda)$, where $m$ is the m'th spectrum observed.

%\clearpage

\subsection{Applying pointing jitter to stellar spectra}
Using the \texttt{Phoenix}\footnote{http://www.hs.uni-hamburg.de/EN/For/ThA/phoenix/index.html} code we generated a stellar spectrum of a sun analogue and proceeded with the following steps:

\begin{enumerate}

\item The input spectrum was trimmed to a wavelength range of 1.0 - 2.5 $\mu$m and re-sampled to a resolution of R = 300 at a central wavelength $\lambda_{c} = 1.75$ $\mu$m. This yields a wavelength coverage of $\delta \lambda = 5.8$ nm per pixel. We denote the spectrum by $F_{star}(\lambda)$, where $\lambda$ is the wavelength.

\item A single stellar absorption line was now selected (wavelength range: 1.19 - 1.21 $\mu$m), and the spectrum consequently trimmed. We denote the trimmed spectrum as $F_{line}(\lambda)$ encompassing $N$ data points in $\lambda$.

\begin{figure}
\center
\includegraphics[width=9cm]{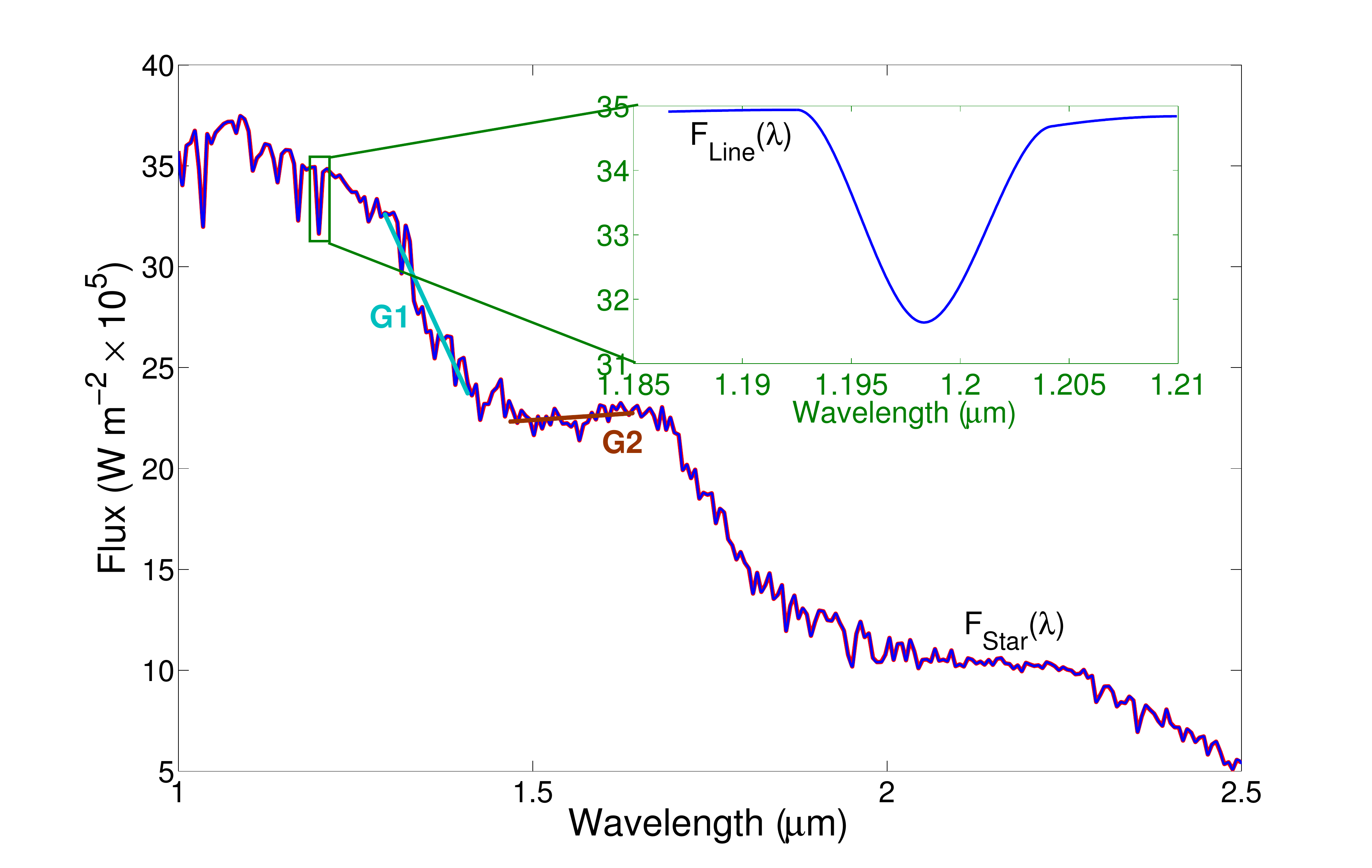}
\caption{Stellar spectrum between 1.0 and 2.5 $\mu$m. Red: down sampled spectrum at R = 300; Blue: interpolated spectrum (see step 4); Inset: spectral line used to fit for the wavelength jitter; G1: flux gradient of stellar `black-body' at 1.29 - 1.40 $\mu$m; G2: flux gradient at 1.46 - 1.64 $\mu$m.}
\label{figspectrum}
\end{figure}

\item Normally distributed random noise was added to the spectrum with the signal-to-noise (SNR) chosen to be the fraction of the maximal line amplitude over the standard deviation of the Gaussian noise component

\begin{equation}
SNR = \frac{max|(F_{line}(\lambda))|}{\sigma_{gauss}}
\end{equation}

\noindent The resulting spectrum, $F_{final}(\lambda)$, is hence $F_{final}(\lambda) = F_{line}(\lambda) + \mathcal{N}(\sigma,\mu)$, where $\mathcal{N}(\sigma,\mu)$ is the Gaussian noise with $\sigma$ variance and $\mu = 0$ mean. We chose a SNR of 500 for this study.

\item The spectrum $F_{final}(\lambda)$ was now interpolated by a factor $\mathcal{S}$ using a `piecewise cubic Hermite interpolation' \citep{press07} which assures the best fit to the original time series. Here we chose $\mathcal{S} = 100$. This step is required to numerically implement sub-pixel drifts and does not impair or bias the results.

\item Steps iii \& iv were repeated $M$ times to create the $M\times N$ dimensional matrix $\bf{X}$, where $N$ is the number of points in $\lambda$ and $M$ was taken to be 100. One can think of $M$ being the time axis containing spectra $F_{final}(1, \lambda), F_{final}(2, \lambda) \dots F_{final}(M,\lambda)$, and $N$ being the spectral axis of the matrix $\bf{X}$.

\item Each spectrum in $\bf{X}$ was now shifted by $\Delta_{m}(\lambda)$ along the wavelength axis randomly towards either the blue or red part of the spectrum.

\end{enumerate}

\subsection{Fitting `jittered' spectra}

\begin{enumerate}

\item Now, each spectrum $F_{final}(m, \lambda)$ was fitted with a Voigt profile using a Levenberg-Marquardt minimisation algorithm \citep{press07} and the resulting centroids, $C_{m}(\lambda)$, were recorded.

\item The recorded centroids were subtracted from the motion jitter to give the fitting residual \\$R_{m}(\lambda)~=~\sqrt{ (\Delta_{m}(\lambda)-C_{m}(\lambda))^{2}} $ and the standard deviation of the residual $\sigma_{R}$, which was recorded.

\item Steps i to ii were repeated 100 times to obtain the sampled probability distribution of P($\sigma_{R}$). 

\end{enumerate}

\subsection{Translating fitting residuals to total flux error}

 We can now use the fitting residual $R_{m}(\lambda)$ to calculate the observed flux, $F_{obs}(m,\lambda)$, per wavelength range $\delta \lambda$ and time step $m$:

\begin{equation}
F_{obs}(m,\lambda) = \int \limits_{\lambda+R_{m}(\lambda)}^{\lambda+\delta \lambda + R_{m}(\lambda)} F_{final}(m, \lambda) \text{d}\lambda
\end{equation}

\noindent above we can see that $R_{m}(\lambda)$ constitutes a change in the integration interval of each pixel with respect to the stellar spectrum. For the case where $\delta \lambda \rightarrow 0$, we can calculate $F_{obs}(m,\lambda)$ using the linear approximation

\begin{equation}
F_{obs}(m,\lambda) =G (R_{m}(\lambda) - \bar{R}_{m}(\lambda)) + F_{final}(m, \lambda)
\end{equation} 

\begin{equation}
G = \frac{\text{d}F_{star}(\lambda)}{\text{d} \lambda}
\end{equation}

\noindent where $\bar{R}_{m}(\lambda)$ is the mean of the fitting residual and was subtracted to account for equal amounts of positive and negative drifts, and $G$ is the local gradient of the stellar spectrum. We can now express the resulting flux error due to residual drifts along the spectral direction as 

\begin{equation}
F_{err}(m,\lambda) = G (R_{m}(\lambda) - \bar{R}_{m}(\lambda))
\label{equfluxerr}
\end{equation}

In section~\ref{sec:results} we show the residual flux error $F_{err}$ for the gradients G1 and G2 shown in figure~\ref{figspectrum}.

\section{Spatial Jitter}
\label{sec:spatial}

\begin{figure}
\center
\includegraphics[width=9cm]{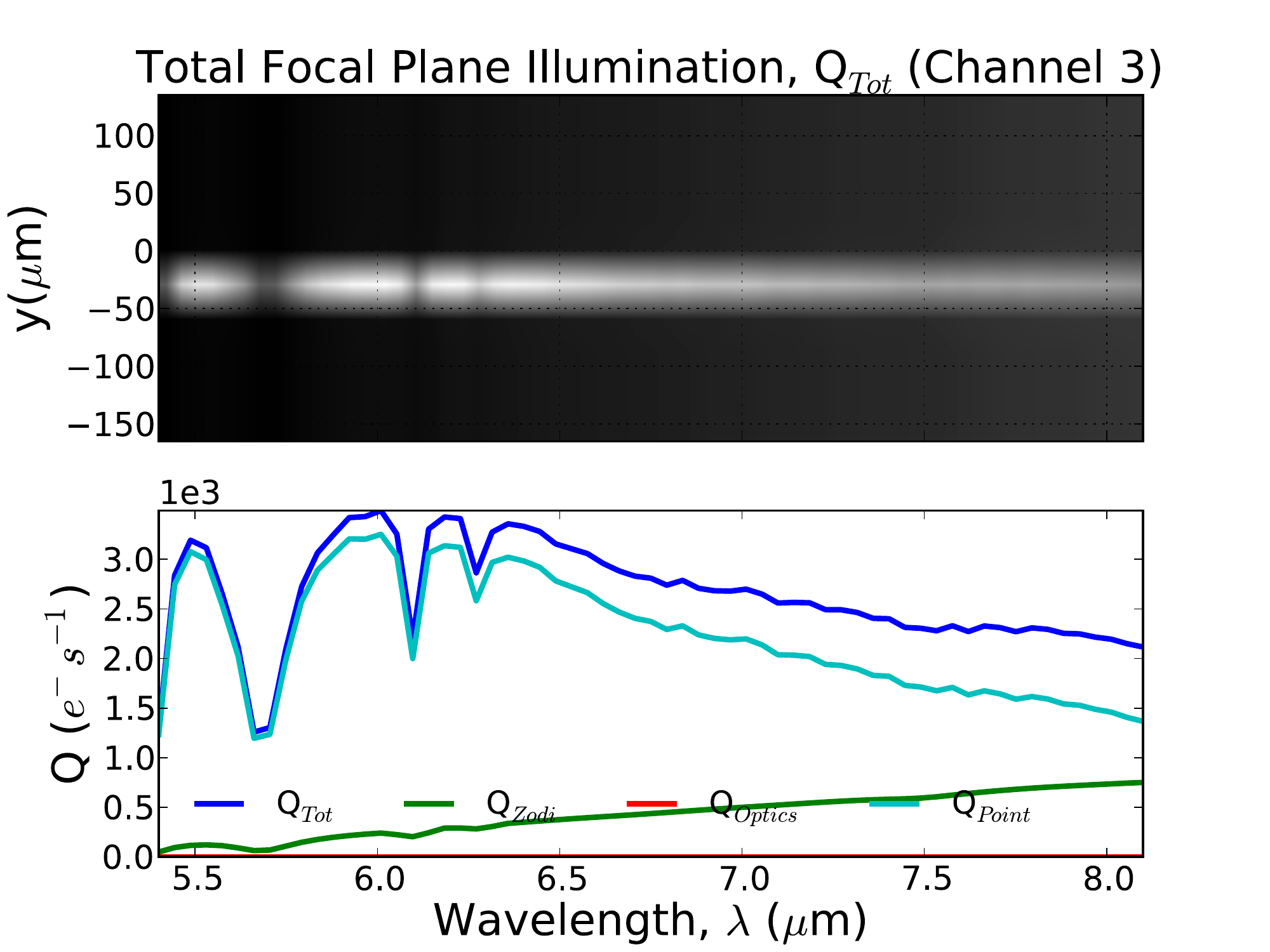}
\caption{Snapshot of \echosim~ output for the MWIR detector. Top: Focal plane illumination including the science spectrum, instrumental and astrophysical backgrounds. Bottom: Plot of the total and individually contributing fluxes. }
\label{spatial1}
\end{figure}

We have investigated the effect of spatial jitter as well as thermal variability. Whereas inter/intra pixel variations can largely be ignored in the case of spectral jitter, the detector pixel responses and variations can play a significant role in the computation of the spatial jitter which is caused by the movement of the spectral PSF along the spatial direction of the detector. \echosim~fully simulates the effect of inter and intra pixel variations on a  spatially resolved detector grid. We here outline the spatial jitter calculations but refer the reader to Pascale et al. (in prep.) for a more detailed discussion.
The dispersed signals are sampled by each detector assuming a wavelength-depended and instrument specific PSF which is convolved with the intra-pixel dependent response of the detector.
 Figure~\ref{spatial1} is a snapshot of a standard diagnostic diagram returned by \echosim. The top panel shows the focal plane illumination of the detector array. The bottom panel shows the individual flux contributions: astrophysical ($Q_{point}$), zodiacal light ($Q_{zodi}$), instrument thermal emissions ($Q_{optics}$), and their combined total ($Q_{tot}$). 

We can describe the focal plane illumination of a detector with a monochromatic point source using the diffraction PSF pattern (Marc Ferlet priv. com.; Pascale et al. in prep.)

\begin{equation}
p(x,y,\lambda) = \frac{1}{2\pi \sigma_{x} \sigma_{y}} e^{\frac{-y^{2}}{2\sigma_{y}^{2}}} e^{-\frac{[x-x_{0}(\lambda)]^{2}}{2\sigma_{x}^{2}}}
\end{equation}

\begin{equation}
\sigma_{x} = \frac{F_{\#}\lambda}{\pi} \sqrt{2/K_{x}} ~~~~~ \sigma_{y} = \frac{F_{\#}\lambda}{\pi} \sqrt{2/K_{y}}
\end{equation}

\noindent where $x$ and $y$ are detector coordinates along the spectral and spatial axes respectively, $K_{x}$ and $K_{y}$ are parameters accounting for spatial aberrations (we assume no aberrations $K_{x} = K_{y}$) and $F_{\#}$ is the ratio between the effective focal length and the effective telescope diameter (see table~\ref{inputtable}).

The one dimensional PSF for the spatial direction can now be written as

\begin{equation}
p_{y}(y,\lambda) = \frac{1}{\sigma_{y} \sqrt{2\pi}} e^{ - \frac{y^{2}}{2\sigma_{y}} }.
\end{equation}

\noindent \citet{barron07} studied the intra-pixel response of IR detectors, and their best-fit model is analytically implemented in \echosim. The cross section of the response along the spatial axis of the detector array is hence given by:

\begin{align}
F(y) &= arctan \left \{ tanh \left [ \frac{1}{2l_{d}} \left ( y + \frac{\Delta_{pix}}{2} \right ) \right ] \right \} \\\nonumber
&- arctan \left \{ tanh \left [ \frac{1}{2l_{d}} \left ( y- \frac{\Delta_{pix}}{2} \right ) \right ] \right \}
\end{align}

\noindent where l$_{d}$ is the diffusion length. The sampled PSF along the spatial axis is then given by the convolution

\begin{equation}
p_{s}(y,\lambda) = p_{y}(y,\lambda) \times F(y)
\end{equation}

\noindent Finally the signal is sampled by the detector. 
The detector response is given by the convolution of the point source flux of the star-planet system, $Q_{point}(\lambda, t)$,  with the spectrally dispersed PSFs

\begin{equation}
Q_{point}(i,j,t) = QE(\lambda) Q_{point} (\lambda,t) \times p_{s}(y,\lambda)
\end{equation}

\noindent where $i$ and $j$ are the detector pixel indices. Note that optical efficiencies don't explicitly appear in this equation as $Q_{point}(\lambda, t)$ already accounts the throughput budget, with the exception of the quantum efficiency, $QE(\lambda)$ (Pascale et al., in prep.).  

\begin{equation}
\sigma_{p}(i,j,t) \simeq \sigma_{p} \frac{f_{eff}}{p_{s}} \frac{\partial p_{s}}{\partial y} Q_{point}
\end{equation}

 \noindent where $f_{eff}$ is the plate scale (in $\mu$m rad$^{-1}$) and $\sigma_{p}$ is the pointing jitter (in radians per second) randomly sampled from the pointing distribution $P(\Delta_{Her})$. 
Variations in photometric errors are estimated by consecutive runs for a range of pointing jitter amplitudes from zero to 200 milli-arcsec over a ten hour observing window. The observed spectrum is reconstructed using the \echosim~pipeline and the uncertainties on the reconstructed spectra are shown in figures~\ref{jitterresult1} \& \ref{jitterresult2} for two cases where PSFs of different sizes are used. 

Given the current uncertainty over the exact nature of inter and intra-pixel variations we have not attempted to de-correlate the pointing jitter in post processing \citep[e.g.][]{swain08,burke10,crouzet12} which would reduce the uncertainties reported in figures~\ref{jitterresult1} \& \ref{jitterresult2}. We must hence consider this analysis as a conservative worst case scenario.

\section{Thermal Stability}
\label{sec:thermal}

We have also investigated the impact of thermal emissions and fluctuations of the optics on the photometric stability of the reconstructed spectrum. Whilst thermal emissions are not directly a problem to photometric stability (source of white noise), they pose constraints on the temperature fluctuations allowed over the time span of an exoplanetary transit/eclipse in the reddest wavelengths. Here varying thermal emissions may produce detector counts that are equivalent or exceed the signal amplitude expected from a planetary eclipse. We calculate the thermal contribution using

\begin{equation}
Q_{thermal}(i,j,t) = \frac{\pi}{4}\frac{\Delta^{2}_{p}}{f^{2}_{\#}} \int \limits^{\lambda_{i}+\frac{1}{2}L\Delta_{p}/LD}_{\lambda_{i}+\frac{1}{2}L\Delta_{p}/LD} BB_{\lambda,T}\text{d}\lambda 
\end{equation}

\noindent where $L$ is the image size of the spectrometer slit in number of pixels, $LD$ the lateral dispersion in mm, $\Delta_{p}$ is the pixel size, $f_{\#}$ is the f-number and $BB$ is the Planck function for a given temperature $T$ and wavelength $\lambda$. \echosim~ calculates the thermal emission of the instrument optics as well as the primary, secondary and tertiary mirrors separately and coherently propagates the resulting emission to detector counts. We have explored a temperature regime ranging from 40 - 60K with a grid size of 0.2K for both, the optical bench and the mirror temperatures. This provides us with an absolute scale of the thermal contributions, see results in section~\ref{sec:results}. We have now calculated the expected signal strength of a secondary eclipse of a HD189733b like hot-Jupiter, $F_{planet}$, and for a given base temperature, $T_{0}$, calculated the temperature variation, $\Delta T$, required to produce a detector signal of the same strength as $F_{planet}$ for a given wavelength $\lambda$:

\begin{equation}
\Delta F_{planet}(\lambda) = f_{\lambda}(T_{0}+\Delta T) - f_{\lambda}(T_{0})
\end{equation}

\noindent where $f_{\lambda}(T)$ is the functional form of the thermal emission/temperature relation for a given wavelength seen in figures~\ref{thermalresults1}~\&~\ref{thermalresults2}. This can either be numerically approximated by interpolation to a fine enough grid or by fitting a high-order polynomial function to the data.

\section{Testcase}

We have now calculated the spectral and spatial jitter contributions for the hot-Jupiter HD189733b \citep{torres08}. We currently have insufficient information on the expected thermal stability in the longest wavelength range so we do not include thermal stability in our test-case calculations. This said, as seen in section~\ref{sec:results}, the overall thermal stability is not a decisive factor for wavelengths shorter than $\sim$14$\mu$m.

For our simulations we use realistic \texttt{Phoenix} stellar models for the host stars. The planetary emission model is taken from \citet{tessenyi12}. \echosim~fully solves the dynamical star-planet system and computes the time resolved spectra by model fitting the generated time series. 

We have now computed the emission spectra of HD189733b with the full noise contribution (shot noise, read noise, spatial/spectral jitter) as well as for the spatial jitter and spectral jitter only. This allows for the direct comparability of the noise contributions.

\section{Results}
\label{sec:results}

\begin{figure}
\includegraphics[width=8cm]{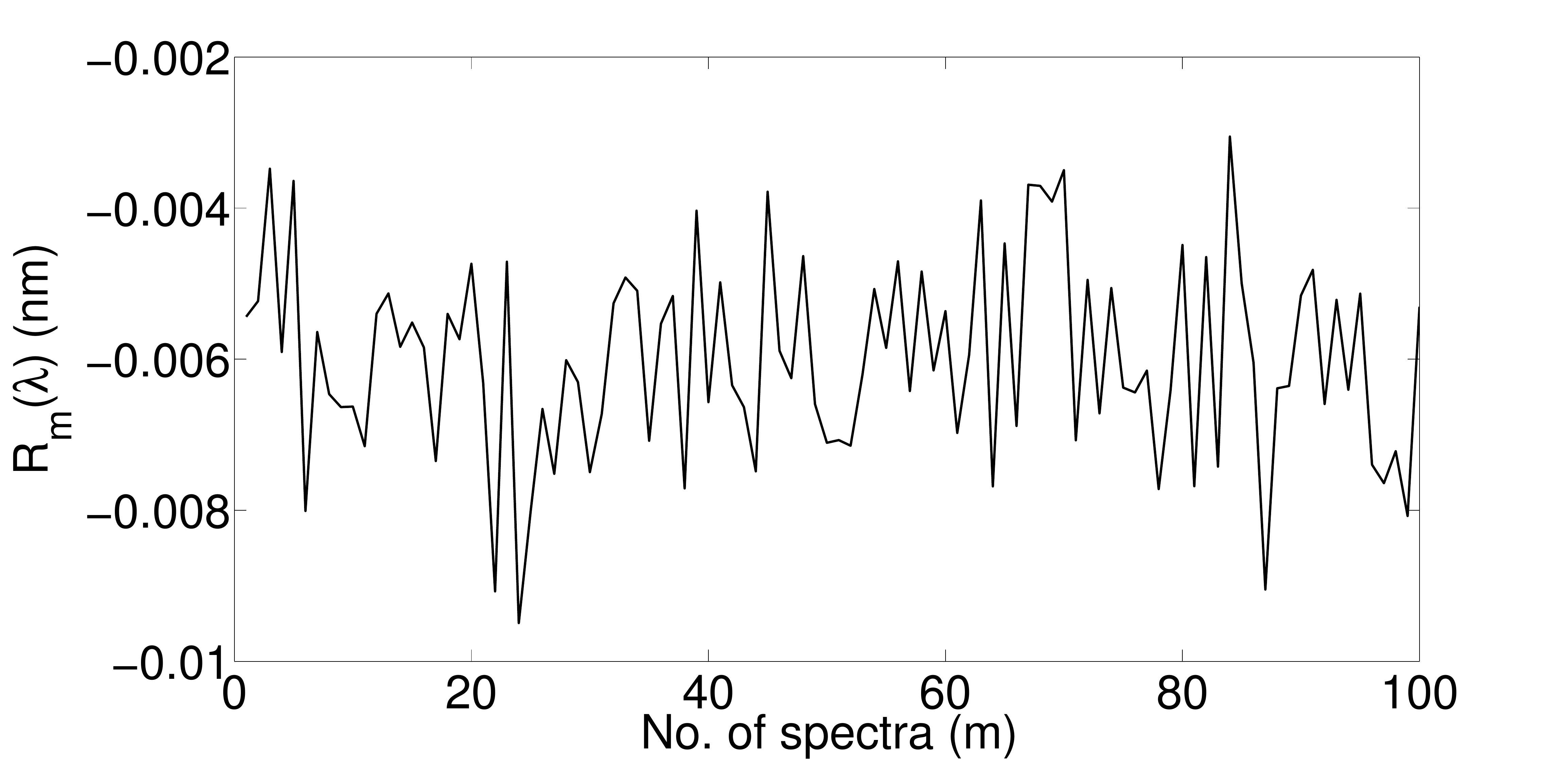}
\caption{Residual centroid fitting error}
\label{figresid}
\end{figure}

\begin{figure}
\includegraphics[width=8cm]{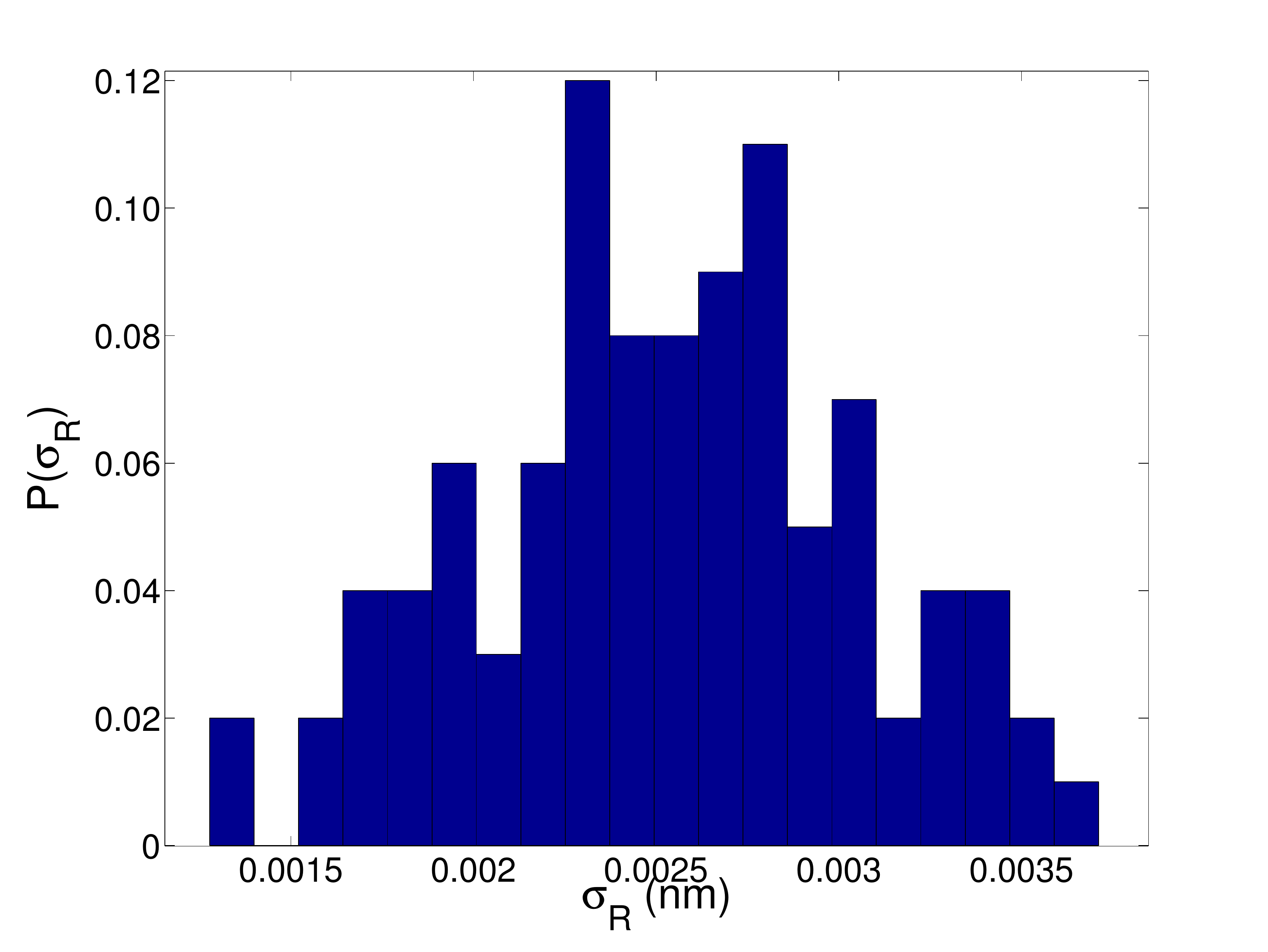}
\caption{Probability distribution of fitting residual standard deviation $\sigma_{R}$ }
\label{fighist}
\end{figure}

\begin{figure}
\includegraphics[width=8cm]{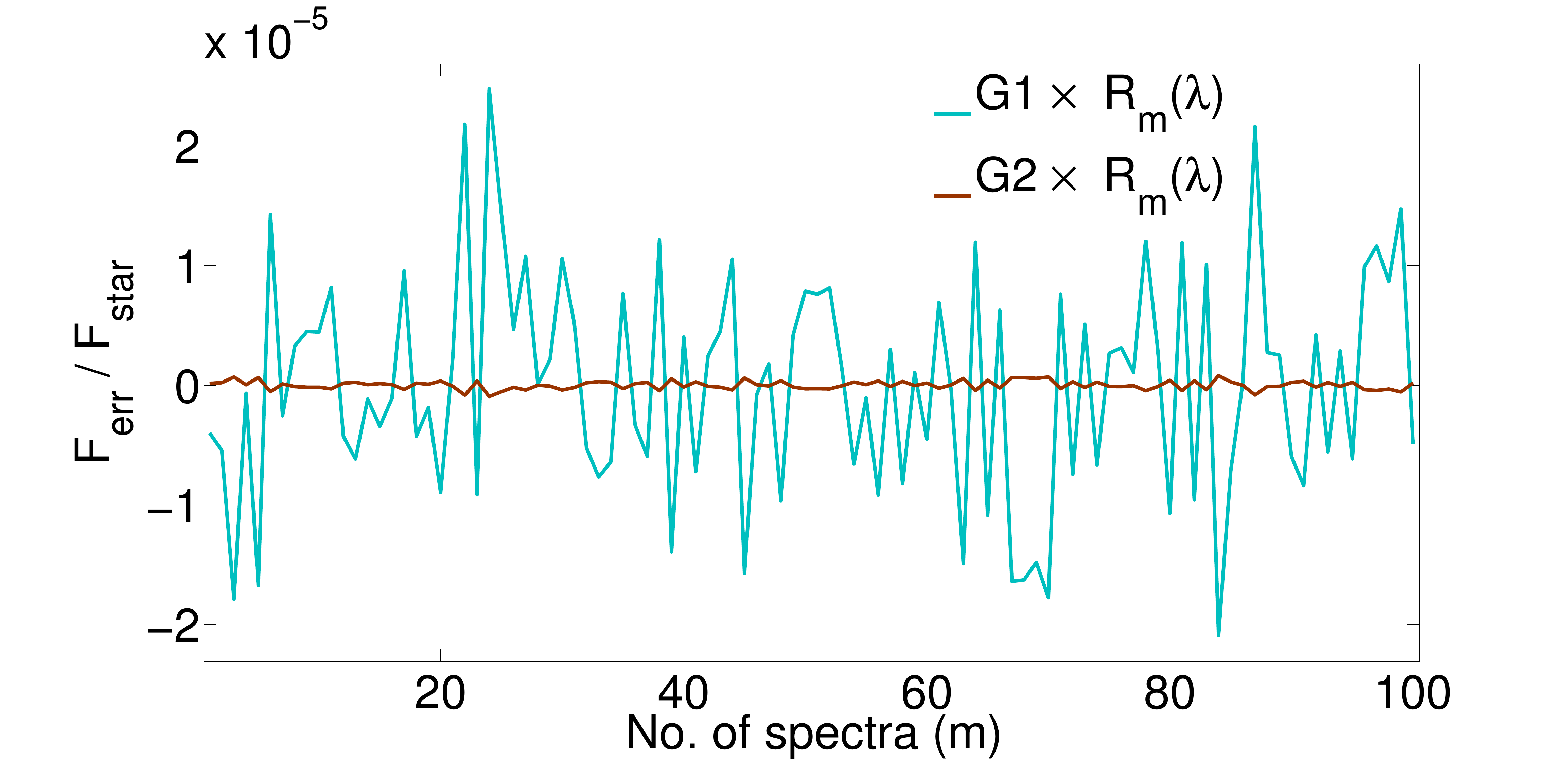}
\caption{Fraction of residual fitting flux $F_{err}$ over the total flux of the star $F_{star}$ for both spectral flux gradients shown in figure~\ref{figspectrum}. G1 = -760 and G2 = 24 $Wm^{-2}nm^{-1}$ .}
\label{figfinalerr}
\end{figure}

\begin{figure}
\includegraphics[width=8cm]{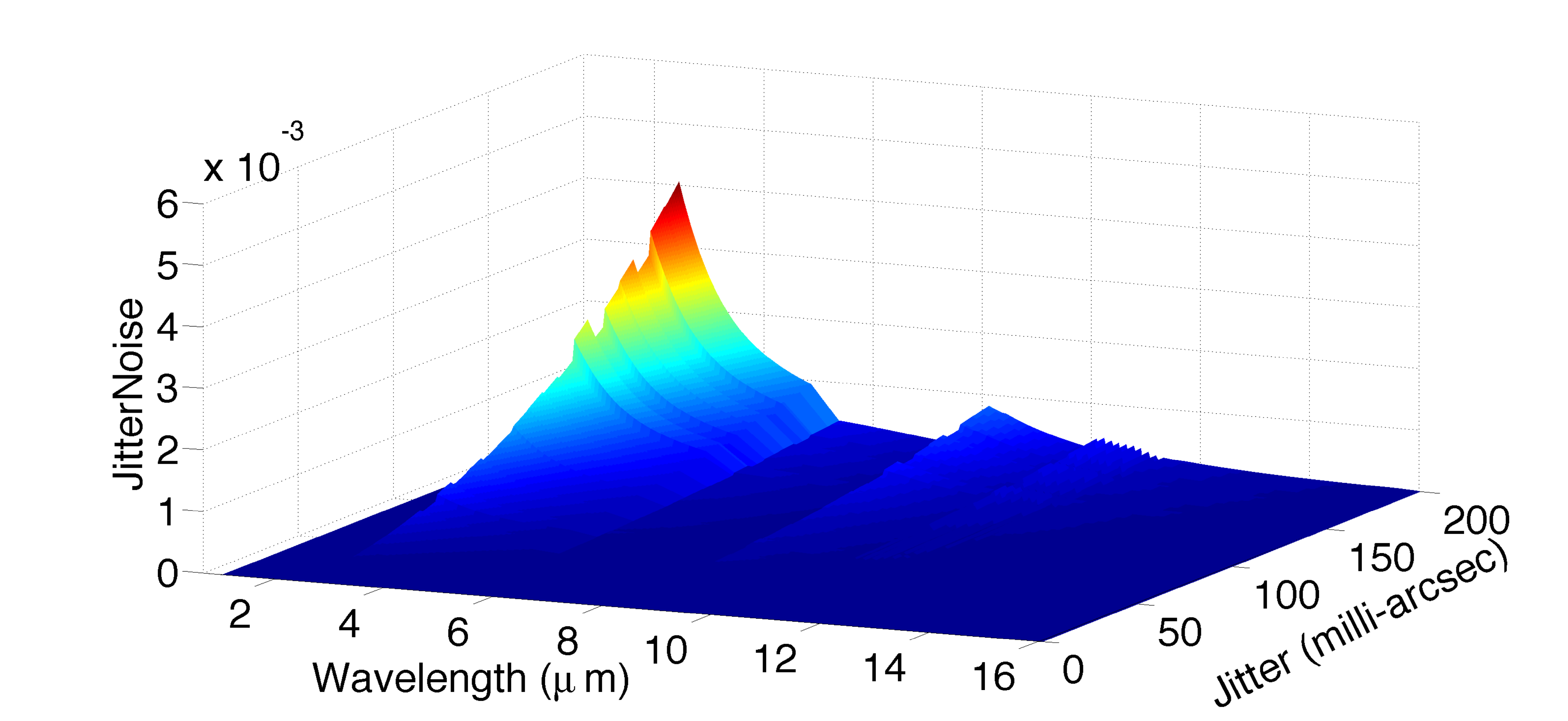}
\caption{Simulation using \echosim~showing spatial jitter noise as fraction of total stellar flux, studied as a function of spectral wavelength and pointing jitter amplitude ranging from 0 Ð 200 milli-arcseconds.  A PSF FWHM is assumed of 0.7 and 0.5 of the detector-pixel size. Figure~\ref{jitterresult2} shows the same simulation with a PSF twice this size. }
\label{jitterresult1}
\end{figure}

\begin{figure}
\includegraphics[width=8cm]{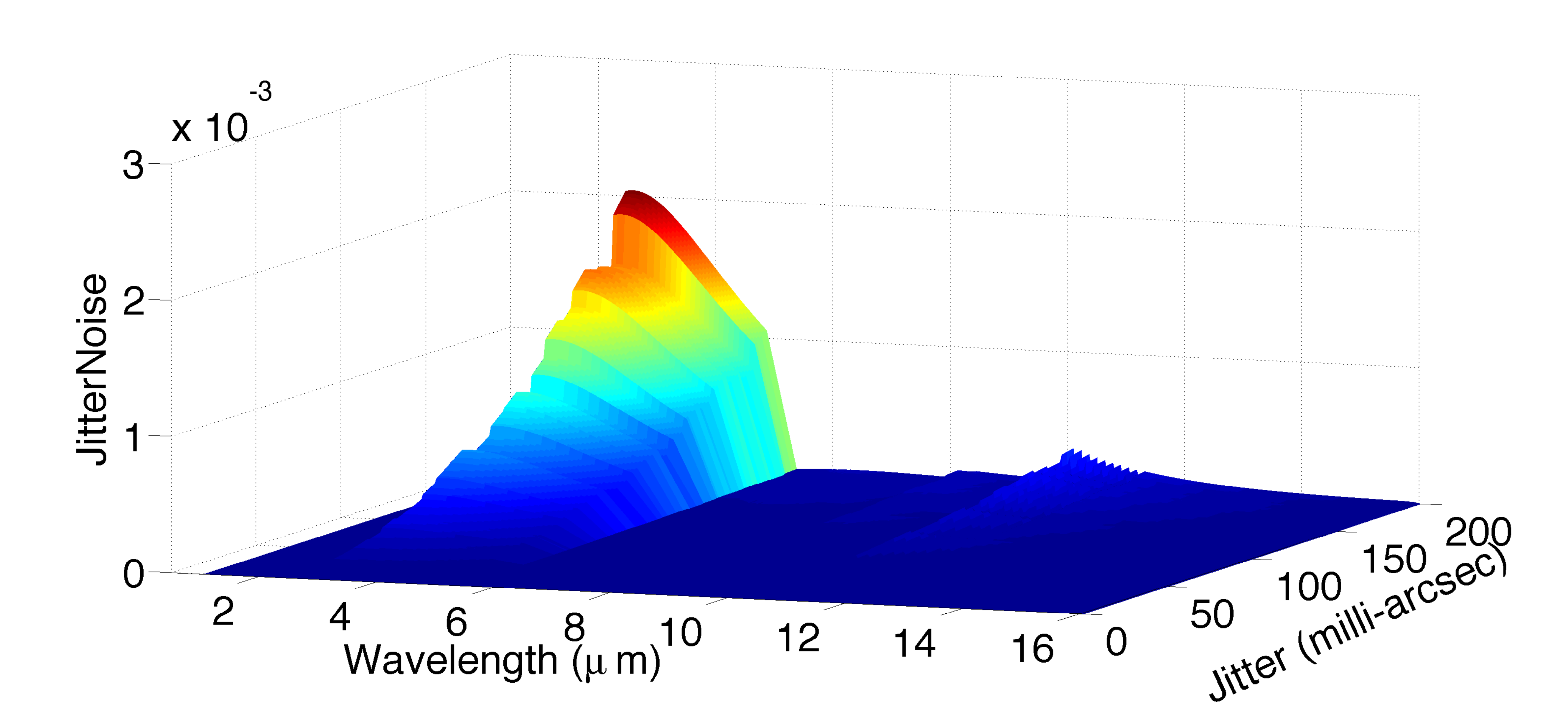}
\caption{Simulation using \echosim~showing spatial jitter noise as fraction of total stellar flux, studied as a function of spectral wavelength and pointing jitter amplitude ranging from 0 Ð 200 milli-arcseconds.  A PSF FWHM is assumed of 1.4 and 1.0 of the detector-pixel size and twice the detector pixel size as stated in table~\ref{inputtable}. }
\label{jitterresult2}
\end{figure}

\begin{figure}
\includegraphics[width=8cm]{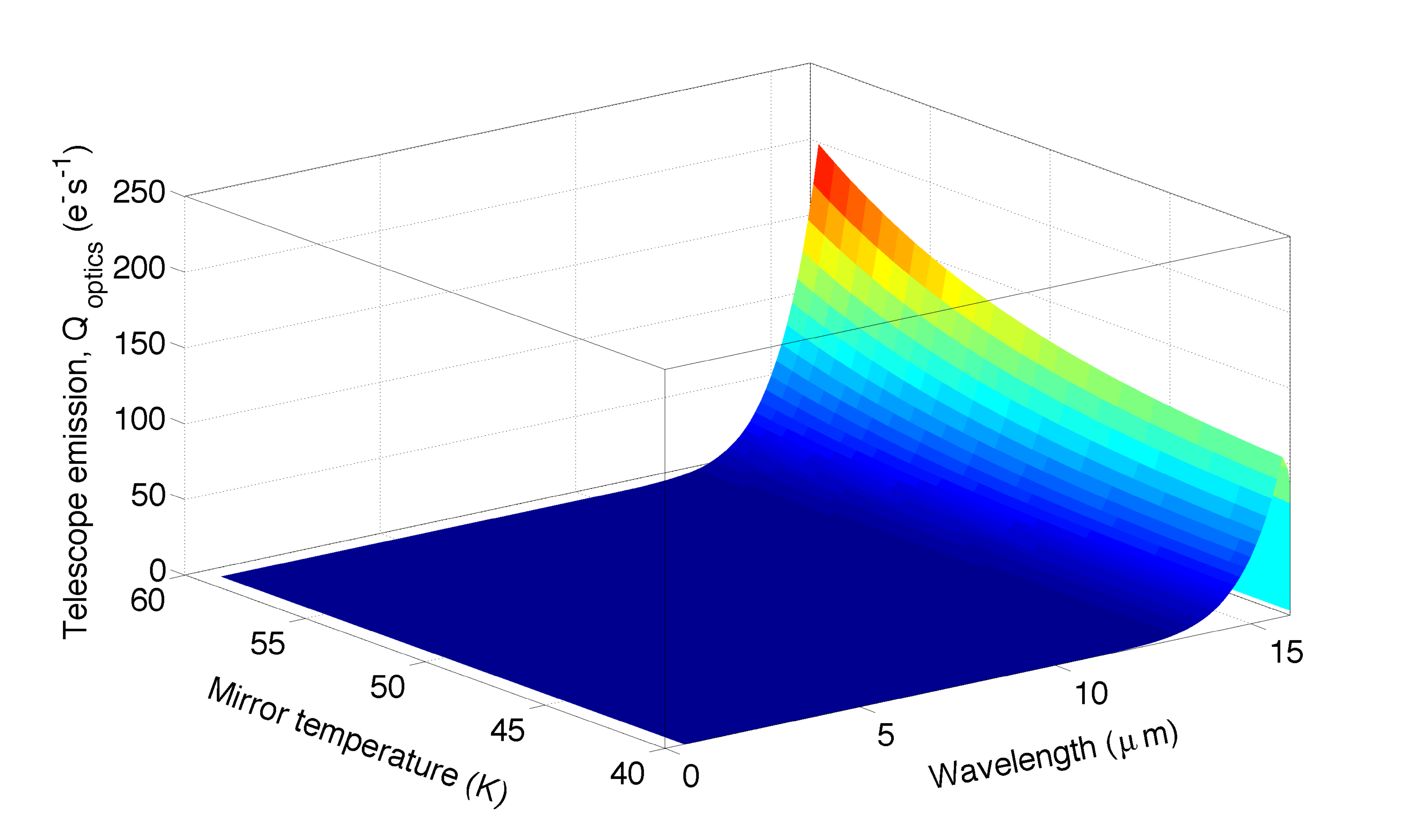}
\caption{Detector counts (in e-/s) due to the telescope emission, as a function of temperature and wavelength.}
\label{thermalresults1}
\end{figure}

\begin{figure}
\includegraphics[width=8cm]{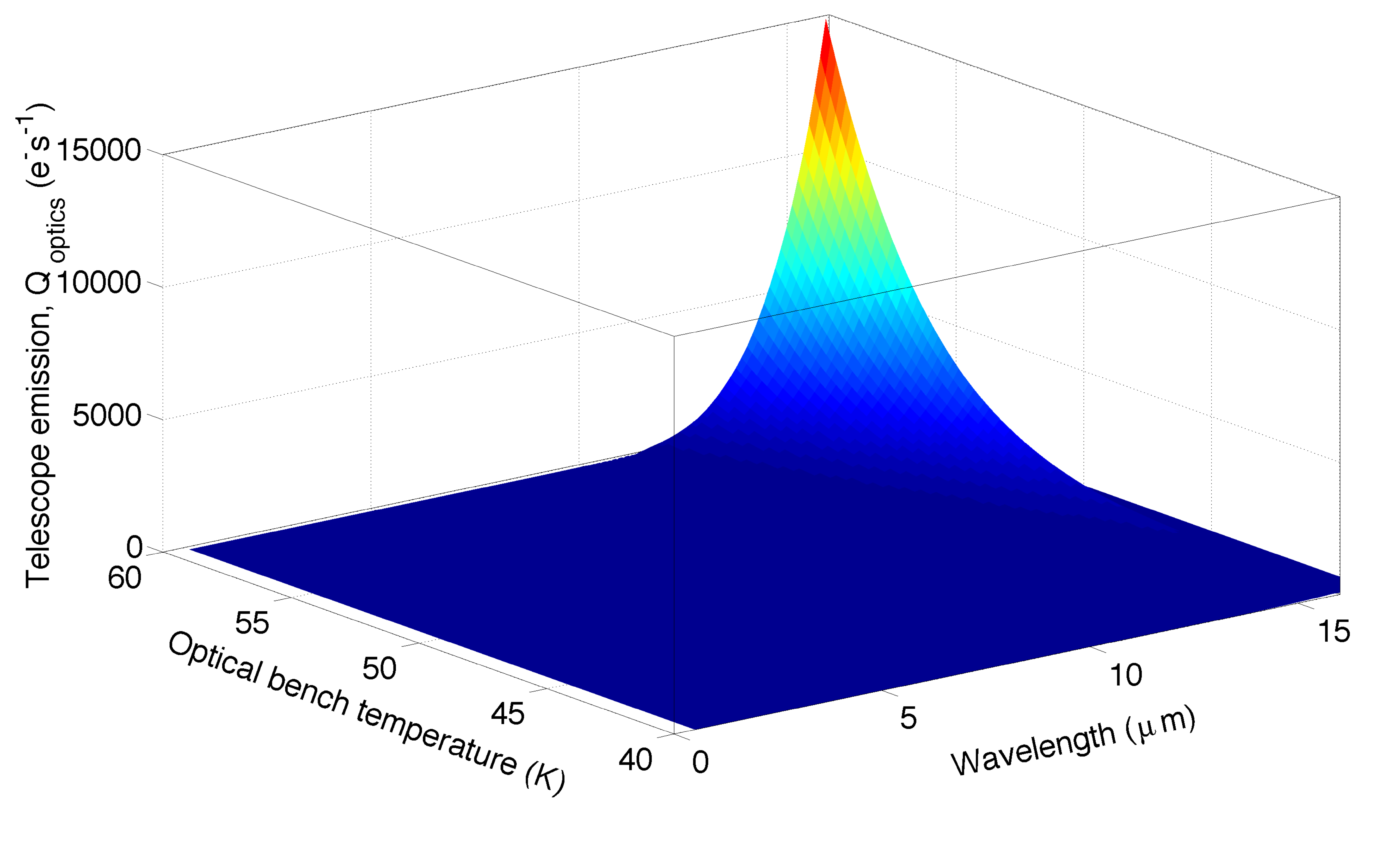}
\caption{Detector counts (in e-/s) due to the optical bench emission, as a function of temperature and wavelength.}
\label{thermalresults2}
\end{figure}

\begin{figure}
\includegraphics[width=8cm]{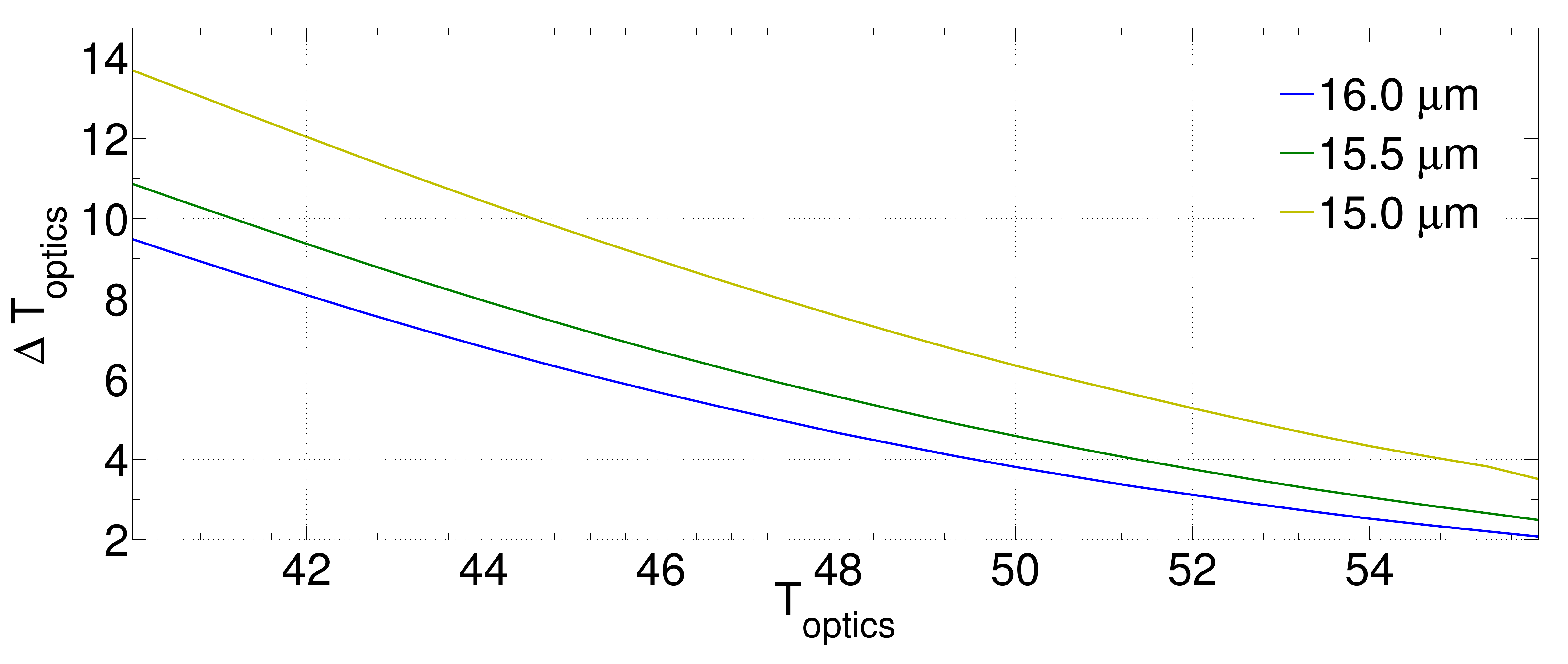}
\caption{The temperature fluctuation required ($\Delta T_{mirror}$) for a given mirror temperature to mimic the variation of counts observed by a secondary eclipse of a HD189733b like hot-Jupiter}
\label{thermalresults3}
\end{figure}

\begin{figure}
\includegraphics[width=8cm]{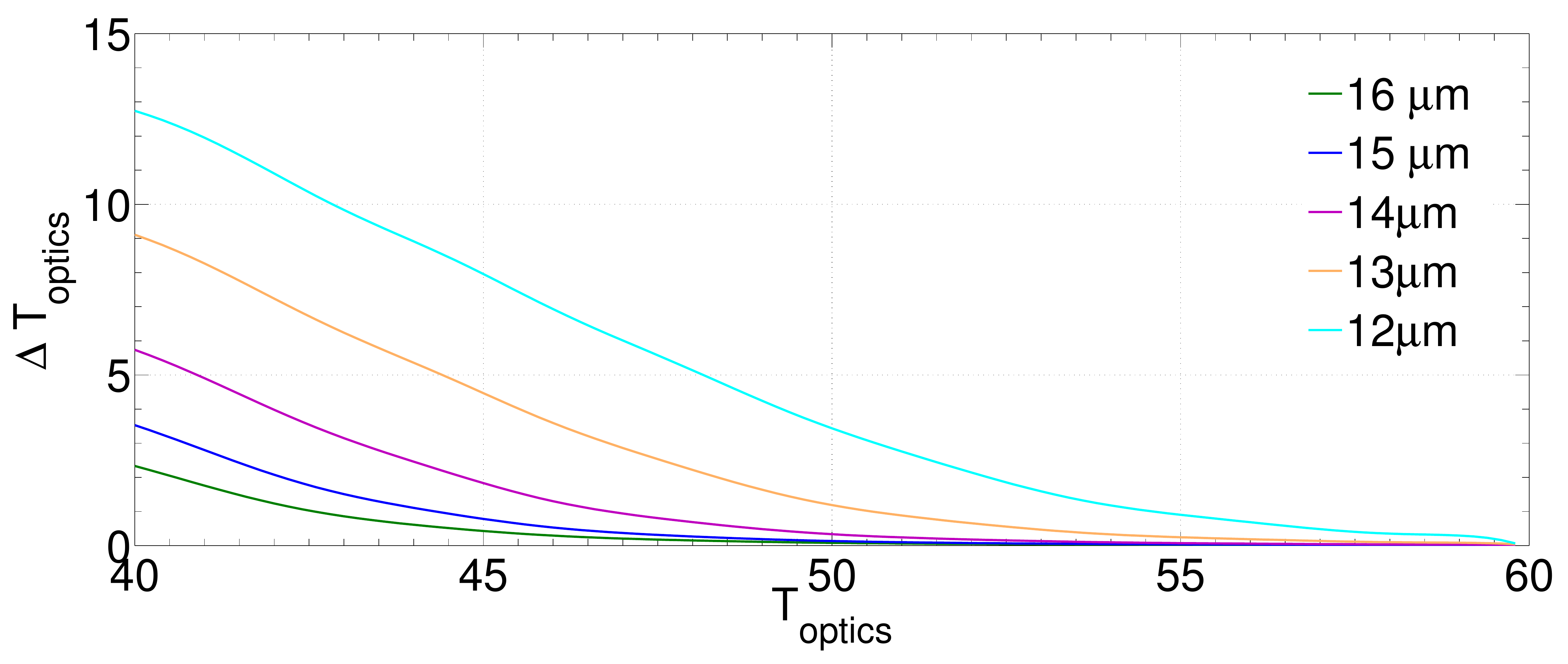}
\caption{The temperature fluctuation required ($\Delta T_{optics}$) for a given optical bench temperature to mimic the variation of counts observed by a secondary eclipse of a HD189733b like hot-Jupiter}
\label{thermalresults4}
\end{figure}

\begin{figure}
\includegraphics[width=9cm]{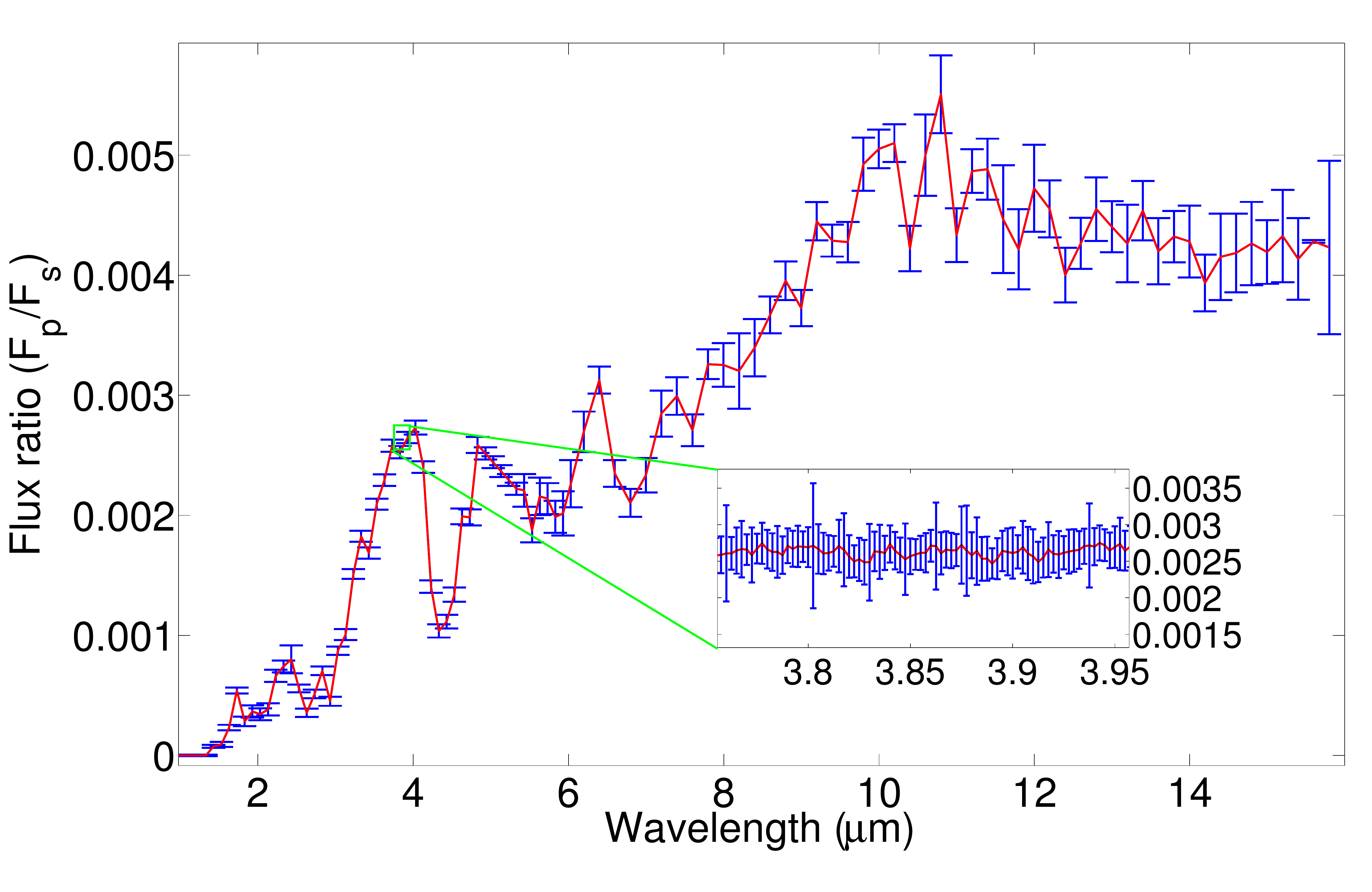}
\caption{Observations of a single secondary eclipse observation of HD189733b binned to 0.1$\mu$m bins. Inset: native resolution of the instrument. }
\label{testcase1}
\end{figure}

\begin{figure}
\includegraphics[width=9cm]{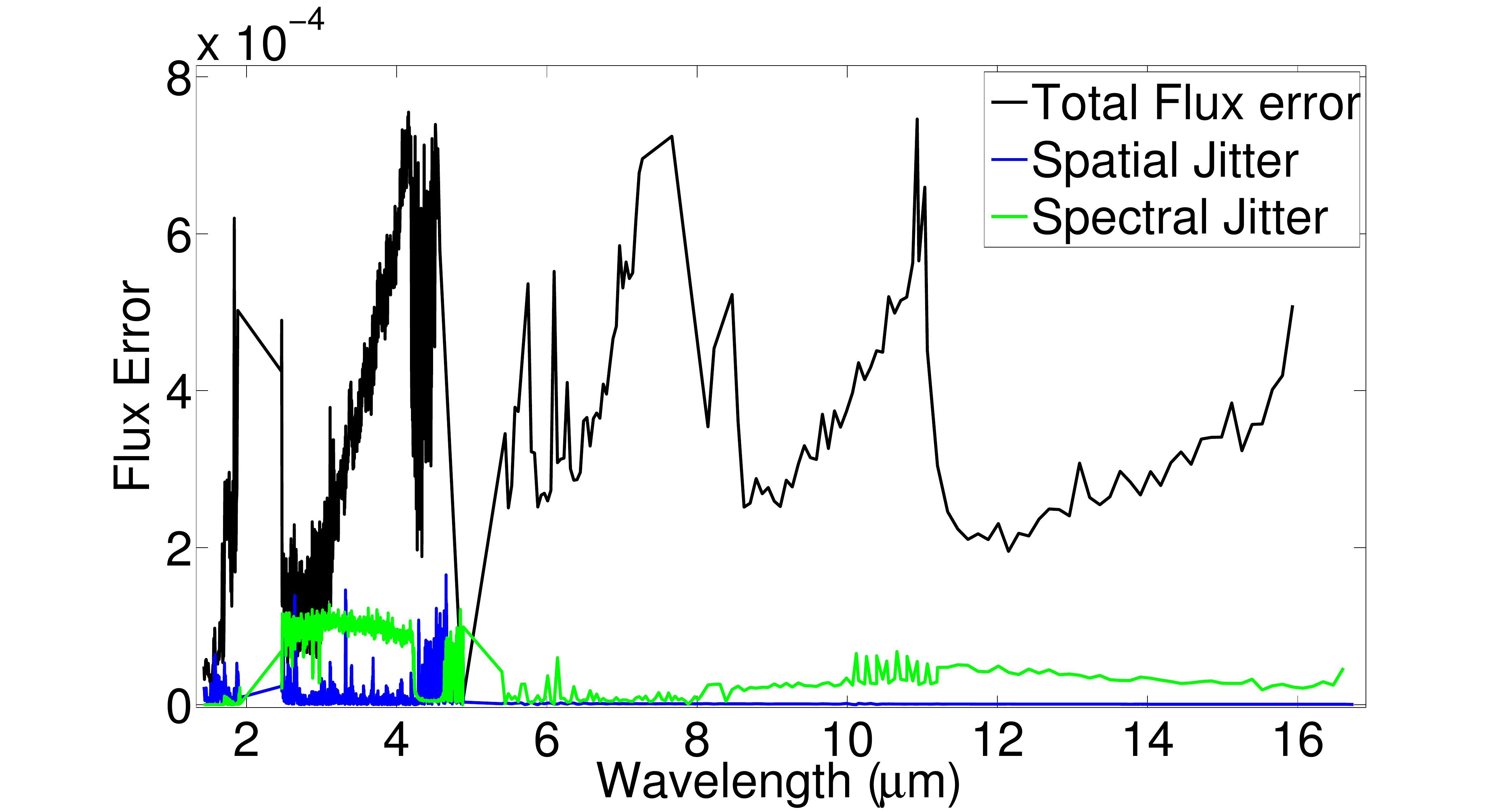}
\caption{Error budget per wavelength in terms of planetary contrast ratio of  HD189733b for spectrum in figure~\ref{testcase1}. Red: total flux error; blue: spectral jitter, green: spatial jitter. }
\label{testcase3}
\end{figure}

As described in section~\ref{sec:spectral} the `jittered' stellar absorption line was fitted using a Voigt profile and the centroids recorded. The residual between the real and fitted spectral shifts are shown in figure~\ref{figresid}. 
The Monte-Carlo analysis of the standard deviation of the fitting residual is shown in figure~\ref{fighist}. As expected, the probability distribution on the parameter $\sigma_{R}$ is largely Gaussian as we randomly sample from the pointing jitter distribution $P(\Delta_{Echo})$.  This also shows that the centroid fitting does not introduce biases in the pointing jitter correction. 
The fitting residual can now be translated to a total flux error using equation~\ref{equfluxerr}. Taking the ratio $F_{err}/F_{star}$ we can derive the relative error due to residual jitter, figure~\ref{figfinalerr}. The flux error is dependent on the local stellar flux gradient. From figure~\ref{figspectrum}, we derived two gradients: G1 = -760 $Wm^{-2}nm^{-1}$ and G2 = +24 $Wm^{-2}nm^{-1}$. For these two gradients, figure~\ref{figfinalerr} shows that the relative flux error lies between $10^{-6} \sim 2 \times 10^{-5}$. These values are of course larger for the Wien tail of the stellar black body with spectral jitter errors of $\sim 10^{-4}$ at places (see figures~\ref{testcase1} \& \ref{testcase3}). 
 
Figures~\ref{jitterresult1} \& \ref{jitterresult2} show the simulated spatial jitter contributions for a pointing jitter amplitude range or 0 - 200 milli-arcseconds. As the photometric error resulting from spatial jitter is largely dependent on the PSF of the instrument and the pixel dimensions, we have calculated the photometric error for a PSF FWHM of 0.7 and 0.5 of the detector-pixel size (figure~\ref{jitterresult1}) and that for double these values (figure~\ref{jitterresult2}). As seen in the figures the spatial jitter is of the order of $\sim 10^{-5} - 10^{-4}$ but significantly higher for the spectral ranges of the NIR instrument  (2 - 5 $\mu$m). This is to be expected as the SWIR instrument features a smaller pixel size.

Figures~\ref{thermalresults1} \& \ref{thermalresults2} show the thermal contribution of the telescope (mirrors and optics) and the optical bench temperatures respectively. This can largely be regarded as negligible for wavelengths shortward of $\sim$14$\mu$m but significant at longer wavelengths and temperatures above 50K. As described in section~\ref{sec:thermal}, this thermal emission was translated into a temperature tolerance, $\Delta T_{optics}$, showing the temperature change necessary to mimic the transit depth of a secondary eclipse feature of hot-Jupiter HD189733b at a given wavelength. These tolerances are shown in figures~\ref{thermalresults3} \& \ref{thermalresults4} for the telescope and optical bench temperatures respectively. We find that when the telescope and optical bench are at 45K, temperature fluctuations need to be below 6K for the telescope and 500mK for the bench in order to observe a hot Jupiter. More restrictive constraints would be required when fainter signals are involved. 

Having investigated the pointing jitter contributions for given wavelengths and pointing jitter amplitudes, we show the simulated `observations'  of a single eclipse event of HD189733b in figure~\ref{testcase1}. The blue spectra's error-bars contain the full noise contribution. Figure~\ref{testcase3} shows a breakdown of the individual error contributions. Here red stands for the total noise contribution (including read and shot noise), blue for the spectral jitter as calculated in section~\ref{sec:spectral} and green for the spatial jitter contribution as calculated in section~\ref{sec:spatial}. It can easily be seen that spectral jitter is important at the Wien tail of the stellar black body distribution and less so at the Rayleigh-Jeans tail. The spatial jitter noise depends on the individual detectors with the effect being strongest for the SWIR detector.

\section{Discussion}

One way to reduce spatial jitter noise is to use pixels that are small compared to the PSF. In this case the effect of the spatial jitter is only governed by the intra-pixel response, and using many pixels to sample the PSF will ÒwashoutÓ both the inter and intra response variations. A higher sampling of the PSF using more pixels will also positively impact the spectral jitter as tracking of the stellar lines becomes easier. 
However, this will only be realistic if the noise from the detectors is sufficiently low but will allow a significant amount of de-correlation as the PSF centroid can be ÒtrackedÓ across the spatial dimension of the array. The simulations take into account realistic intra-pixel responses in the spectral band from MWIR and LWIR. 
At shorter wavelength, the simulations currently have to be considered worst case scenarios because the system is not diffraction limited at these wavelength.

Figures~\ref{thermalresults1} \& \ref{thermalresults2} show the detected radiation in electrons/second from the instrument and telescope (separately) for a temperature range of 40 - 60K. Further studies will include additional instrumental effects such as mechanical vibrations, thermo-mechanical distortions, variable detector dark currents (assumed fixed with temperature in this study), detector responsitivity drifts and the effect of cosmic ray impacts. These studies will
look at the extent the effect can be de-correlated from the timelines when additional information is available for data processing, such as temperature sensors monitoring the optics and telescope temperatures. We also plan to provide off-axis detectors to monitor the non stellar backgrounds and therefore provide a means of directly removing the background signals. Such background removal is particularly important for fainter sources. Whereas HD189733b is photon limited, observations of faint super-earths will likely be background limited and thermal and zodiacal light emissions need to be carefully accounted for.

\section{Conclusion}

In this paper we present the methodology used for a photometric stability analysis of the \echo~mission and asses the photometric stability given its current `Phase-A' design specifications. We describe how spectral and spatial jitter due to space-craft pointing uncertainties are propagated to an uncertainty on the exoplanetary spectrum measured by \echo. We furthermore investigate tolerances on the thermal stability of the space-craft's optical path. The photometric stability error budget was estimated for a simulated secondary eclipse observations of the hot-Jupiter HD189733b. As the instrument parameters are not set in stone as of date, we have throughout considered the `worst-case' assumptions only and photometric stability errors may significantly decrease as the instrument definition phase proceeds.

\section*{Acknowledgments}

This work is supported by STFC, NERC, UKSA, UCL and the Royal Society.

\label{lastpage}

\end{document}